\documentclass[pra,twocolumn,preprintnumbers,amsmath,amssymb,nofootinbib]{revtex4}
\usepackage[T1]{fontenc}
\usepackage[latin1]{inputenc}
\usepackage{amsmath,amsfonts, graphicx,epsfig, exscale, relsize,pstricks,color}
\usepackage{slashed}
\usepackage{bbm,bm}
\usepackage{hyperref}
\usepackage[active]{srcltx}

\usepackage{color}
\definecolor{holger}{rgb}{0,0.4,0.7}
\definecolor{comment}{rgb}{0.9,0,0}
\definecolor{alexej}{rgb}{0.5,0.5,0}

\newcommand{\pt}{\mathcal{T}}
\newcommand{\nL}{{n_{\text{L}}}}
\newcommand{\bgam}{\boldsymbol{\gamma}}

\newcommand{\Eqref}[1]{Eq.~\eqref{#1}}

\begin{document}

\title{{Geothermal Casimir phenomena for the sphere-plate and  cylinder-plate configurations}}
\author{Alexej Weber${}^1$ and Holger Gies${}^{2}$}

\affiliation{
\mbox{\it ${}^1$Institut f{\"u}r Theoretische Physik, Universit{\"a}t Heidelberg,
Philosophenweg 16, D-69120 Heidelberg, Germany} 
\mbox{\it ${}^2$Theoretisch-Physikalisches Institut, Friedrich-Schiller-Universit{\"a}t Jena,
Max-Wien-Platz 1, D-07743 Jena, Germany}
\mbox{\it E-mail: {a.weber@thphys.uni-heidelberg.de, holger.gies@uni-jena.de}}
}

\begin{abstract}
{  We investigate the nontrivial interplay between geometry and temperature in
  the Casimir effect for the sphere-plate and cylinder-plate configurations.
  At low temperature, the thermal contribution to the Casimir force is
  dominated by this interplay, implying that standard approximation techniques
  such as the PFA are inapplicable even in the limit of small
  surface separation. Thermal fluctuations on scales of the thermal wavelength
  lead to a delocalization of the thermal force density at low temperatures.
  As a consequence, the temperature dependence strongly differs from naive
  expectations. Most prominently, thermal forces can develop non-monotonic
  behavior below a critical temperature. We perform a comprehensive study of
  such geothermal phenomena in these Casimir geometries, using analytical and
  numerical worldline techniques for Dirichlet scalar fluctuations.}
\end{abstract}

\maketitle

\section{Introduction}

The Casimir effect \cite{Casimir:dh}, inspiring many branches of physics
\cite{Bordag:2001qi,Milton:2001yy}, features a decisive geometry dependence:
the fluctuation-induced interaction between test bodies or surfaces depends on
their shape and orientation. This is because the Casimir effect arises from
the fluctuation spectrum in presence of the surfaces relative to the vacuum
fluctuations. The spectral properties in turn are a direct consequence of the
geometry.

This geometry dependence becomes even more pronounced at finite temperature
$T$: thermal fluctuations can predominantly be associated with a
characteristic length scale, the thermal wavelength $\lambda_T \sim \hbar c /
(k_{\text{B}} T)$. Thermal fluctuations contribute to the Casimir force,
whenever the scale set by the thermal wavelength is commensurate with a mode
of the fluctuation spectrum as defined by the geometry. Therefore, thermal
corrections to the zero-temperature Casimir effect generally cannot be
described by universal additive terms or other simple recipes but require a
careful analysis of the interplay between geometry and temperature, as first
anticipated in  \cite{Scardicchio:2005di}.

This ``geothermal'' interplay has first been verified in paradigmatic
perpendicular-plates \cite{Gies:2008zz} or general inclined-plates
configurations \cite{Weber:2009dp}. Further evidence for the experimentally
relevant sphere-plate configuration has been provided recently in
\cite{Canaguier:2009,Gies:2009nn,Bordag:2009dz,Zandi}. Typical low-temperature
dependencies in these open geometries obey power laws with characteristic
exponents that are particular for the geometry. Most importantly, these
power laws disagree with predictions from standard local approximation
techniques such as the proximity force approximation (PFA) \cite{pft1}  -- even in the limit
of vanishing surface separation. This is in contrast to zero-temperature
forces which are often well described by the PFA in this limit \cite{semicl}.

In this work, we perform a comprehensive study of the geometry-temperature
interplay for the sphere-plate and cylinder-plate configuration. We study the
Casimir forces induced by fluctuations of a scalar field obeying Dirichlet
boundary conditions on the surfaces in order to explore the geothermal
interplay in a most transparent fashion. Moreover, we use the worldline
approach to the Casimir effect \cite{Gies:2003cv} which on the one hand
provides for a highly intuitive picture of the fluctuations, and on the other
hand facilitates analytical as well as numerical computations from first
principles \cite{Feynman,Halpern:1977ru,Schmidt:1993rk,Gies:2001zp}.

For instance, the failure of local or additive approximation techniques can
directly be inferred from the temperature dependence of the force density: the
latter tends to delocalize for decreasing temperatures on scales of the
thermal wavelength \cite{Gies:2009nn}. Local approximation techniques may only
be useful at finite temperature if the strict weak-coupling limit is taken
\cite{Milton:2009gk}, or in the high-temperature limit.

In the present work, we analyze the thermal force density distributions,
compute thermal forces for a wide nonperturbative range of parameters, and
determine asymptotic limits. This facilitates a careful comparison with local
approximation techniques, and, most importantly, yields new and unexpected
results for the geometry dependence of thermal forces. For instance, the pure
thermal force, i.e., the thermal contribution to the Casimir force, reveals a
non-monotonic behavior below a critical temperature for the sphere-plate and
cylinder-plate case \cite{Weber:2010kc}: the attractive thermal force can
increase for increasing distances. This anomalous feature is triggered by a
reweighting of relevant fluctuations on the scale of the thermal wavelength --
a phenomenon which becomes transparent within the worldline picture of the
Casimir effect.  Whereas these non-monotonic features already occur for a
simple Dirichlet scalar model, non-monotonicities can also arise from a
competition between TE and TM modes of electromagnetic fluctuations in
configurations with side walls \cite{Rodriguez:2007,Rahi}.

While there are a number of impressive verifications of the zero-temperature
Casimir force \cite{Lamoreaux:1996wh}, a comparison between theory and thermal
force measurements suffers from the interplay between dielectric material
properties and finite temperature \cite{Sernelius}, still {being a subject of
  intense theoretical investigations}
\cite{Mostepanenko:2005qh,Brevik:2006jw,Bimonte:2009nf,Ingold,Intravaia:2009pu,Ellingsen:2010}. In
view of the geothermal interplay, we expect that the full resolution of this
issue requires the comprehensive treatment of geometry, temperature and
material properties, possibly also including edge effects
\cite{Gies:2006xe,Graham:2009zb,Weber:2009dp,Kabat:2010nm,Onofrio}. First
results on the sphere-plate configuration using scattering theory and specific
dielectric models demonstrate this nontrivial interplay
\cite{CanaguierDurand:2009zz,Canaguier:2009,Zandi}. 

As a crucial ingredient for such an analysis, field-theoretical methods for
Casimir phenomena have to be used that can deal with arbitrary Casimir
geometries. In addition to the worldline methods
\cite{Gies:2003cv,Gies:2005ym,Gies:2006bt,Gies:2006cq,Schaden:2009zz} used in this work, a
variety of approaches has been developed in recent years, such as a functional
integral approach \cite{Bordag:1983zk,Emig:2001dx,Emig:2003eq} and scattering
theory \cite{Bulgac:2005ku,Emig:2006uh,Bordag:2006vc,%
  Kenneth:2006vr,Emig:2007cf,Rodrigues:2006ku,Mazzitelli:2006ne,%
  Milton:2007gy,Milton:2008vr}.  An extension of these methods to finite
temperature is usually straightforward and highly worthwhile in view of the
geometry-temperature interplay. 

Our paper is organized as follows: after a brief account of the worldline
approach to the Casimir effect in Sect.~\ref{sec:worldl-appr-casim}, the
sphere-plate and cylinder-plate configurations are studied at zero temperature
in Sect.~\ref{sec:case-t=0}. In addition to making contact with the
literature, we perform the worldline computation directly for the force
instead of the interaction energy. Section \ref{sec:fin-T} contains all our
main results on the finite-temperature case. Our conclusions are summarized in
Sect.~\ref{sec:conclusions}. For reasons of comparison, the proximity-force
approximation for the sphere-plate and cylinder-plate case is worked out in
detail in {appendix \ref{app:PFA}}. In addition to explicit formulas which have not
comprehensively appeared in the literature so far, we relate the PFA to an
approximate treatment of the worldline path integral which helps to understand
the differences between the exact and approximate treatments.

\section{Worldline approach to the Casimir effect}
\label{sec:worldl-appr-casim}

We start with a short reminder of the worldline approach to the Casimir effect for a
massless Dirichlet scalar in 4 dimensions; for details, see \cite{Gies:2003cv,Gies:2006cq,Weber:2009dp}.
For a configuration $\Sigma$ consisting of two rigid objects with
surfaces $\Sigma_1$ and $\Sigma_2$, the worldline representation of the
Casimir interaction energy in $D=4$ dimensional spacetime reads
\begin{align}\label{Int-1}
E_\mathrm{c}=-\frac{1}{32 \pi^{2}} \!\int_{0}^\infty
 \frac{\mathrm{d}\pt}{\pt^{3}} \!\int \!\mathrm{d}^3
x_\mathrm{CM}\left\langle 
\Theta_\Sigma [\mathbf{x}(\tau)] \right\rangle.
\end{align}
The worldline functional $\Theta_\Sigma[\mathbf{x}(\tau)]$ is $1$
if the  worldline $\mathbf{x}(\tau)$ intersects both objects
$\Sigma=\Sigma_1\cup\Sigma_2$, and is zero otherwise. 

The expectation value in \Eqref{Int-1} is taken with respect to an ensemble of
$3$-dimensional closed worldlines with a common center of mass
$\mathbf{x}_{\mathrm{CM}}$ and obeying 
a Gau\ss ian velocity distribution.
For static Casimir configurations the time component cancels out  at zero
temperature.  Equation (\ref{Int-1}) has an
intuitive interpretation: All worldlines intersecting both surfaces violate Dirichlet boundary conditions and are removed from
the ensemble of allowed fluctuations,
contributing to the negative Casimir interaction energy. During the
${\pt}$ integration, the extent of a  worldline is scaled by  $\sqrt{\pt}$ . Large propertimes
$\pt$ correspond to  IR fluctuations, small
$\pt$ to  UV fluctuations.

Introducing finite temperature $T=1/\beta$ by the Matsubara formalism is
  equivalent to compactifying {Euclidean} time on the interval
$[0,\beta]$. 
The Casimir free energy corresponding to
 \Eqref{Int-1} now becomes
\begin{eqnarray}\label{Int-3}
E_\mathrm{c}&=&-\frac{1}{32 \pi^{2}} 
\\
&&\times \! \! \int_{0}^\infty
\frac{\mathrm{d}\pt}{\pt^{3}}\!\!\sum_{n=-\infty}^{\infty} \!\!e^{-\frac{n^2\beta^2}{4
\pt}}\!\! \int \! \mathrm{d}^d x_\mathrm{CM}\left\langle
\Theta_\Sigma [\mathbf{x}(\tau)] \right\rangle.\nonumber
\end{eqnarray}
For numerical purposes, it is convenient to remove the $\pt$ dependence
  from the velocity distribution by the rescaling
\begin{equation}
\bgam(t):= \frac{1}{\sqrt{\pt}} \mathbf x(\pt t)\quad \!\!\!\to\!\!\!\quad
e^{-\frac{1}{4} \int_0^{\pt}   \dot{\mathbf{x}}^2 \mathrm{d}\tau} = e^{-\frac{1}{4}
  \int_0^1  \dot{\bgam}^2 \mathrm{d}t},
\end{equation}
where 
$\dot{\bgam}=\mathrm{d}\bgam(t)/\mathrm{d}t$. 
Now, the $\Theta$ function reads more explicitly
\begin{equation}
\Theta[\mathbf x]\equiv\Theta[\mathbf
x_{\text{CM}}+\sqrt{\pt}\bgam(t)].
\end{equation}
The worldline integrals are evaluated numerically by Monte Carlo methods, i.e,
the path integral is approximated by a sum over a finite ensemble of $\nL$
worldlines.  Each worldline $\bgam(t)$ is furthermore discretized by a finite
set of $N$ points per loop (ppl).  To generate discretized worldlines with
Gau\ss ian velocity distribution the v-loop algorithm was used in this work
\cite{Gies:2003cv,Gies:2005sb}.  In the remainder, we apply the worldline
method to the sphere-plate and cylinder-plate Casimir configurations.

\section{Casimir effect at zero temperature}
\label{sec:case-t=0}

{Let us first study the sphere-plate and cylinder-plate geometries at zero
  temperature. Here, we make contact with earlier results for the Casimir
  effect of a cylinder and sphere above a plate
  \cite{Gies:2001zp,Bulgac:2005ku,Gies:2006bt,Emig:2006uh,Bordag:2006vc,Gies:2006cq}. Moreover,
  we generalize the worldline method to directly compute the Casimir force
  instead of the energy which leads again to significant simplifications
  compared to previous energy calculations.} The method will be generalized to
finite temperature in the next section.

\subsection{Sphere above a plate}
\begin{figure}[t]
\begin{center}
\includegraphics[width=0.95\linewidth]{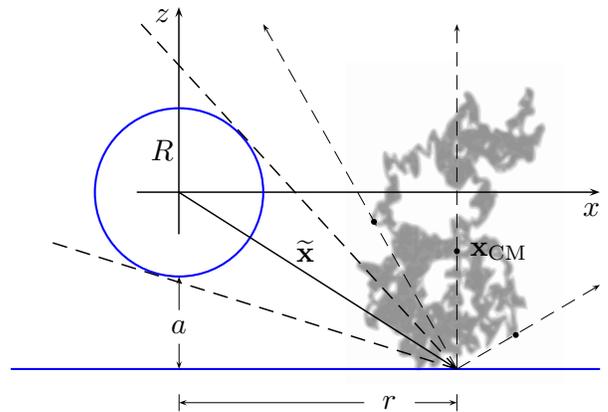}
\end{center}
\caption{Sketch of the sphere-plate configuration. The infinite plate (blue
  line) is in the $z=-(R+a)$ plane. The sphere of radius $R$ is in the
  origin. The center of mass $\mathbf{x}_\mathrm{CM}$ of the worldline is in
  $(r,0,-a-R-\sqrt{\mathcal{T}}\bgam_{z_\mathbf{\min}})$. During the
  propertime integration the worldline always touches the plate, while all its
  points move on rays passing through $(r,0,-a-R)$.  Only points lying inside
  the cone will pass through the sphere.}\label{CEff-SaP-10}
\end{figure}

{We start the configuration of a sphere above a plate. The sphere of
  radius $R$ is centered around the origin $\mathbf{x}=0$. The infinitely
  extended plate lies in the $z=-(a+R)$ plane, where $a$ is the minimal
  distance between both objects, see Fig. \ref{CEff-SaP-10}.  Since the
  configuration has a rotational symmetry with respect to the $z$ axis,} the
  three dimensional $\mathbf{x}_\mathrm{CM}$ integration reduces to a two
  dimensional one.  The Casimir  energy (\ref{Int-1}) reads
\begin{align}\label{CEff-SaP-1}
E_\mathrm{c}=-\frac{1}{16 \pi}\!\int_{0}^\infty  \frac{\mathrm{d}\mathcal{T}}{\mathcal{T}^{3}}
\! \int \! \mathrm{d}r \, \mathrm{d}
z_\mathrm{CM} \,
r\left\langle
\Theta_\Sigma [\mathbf{x}(\tau)] \right\rangle,
\end{align}
where we have switched to cylindrical coordinates
{($r,z_\mathrm{CM}$) with 
$r^2=x_\mathrm{CM}^2+y_\mathrm{CM}^2$.} The $\Theta_\Sigma
[\mathbf{x}(\tau)]$ functional factorizes,
\begin{align}\label{CEff-SaP-2}
\Theta_\Sigma [\mathbf{x}(\tau)] =\Theta_\mathrm{S} [\mathbf{x}_\mathrm{CM}+\sqrt{\mathcal{T}} \bgam]\Theta_\mathrm{P} [\mathbf{x}_\mathrm{CM}+\sqrt{\mathcal{T}} \bgam]. 
\end{align}
Here $\Theta_\mathrm{S}$ and $\Theta_\mathrm{P}$ account for the intersection
of a worldline $\mathbf{x}_\mathrm{CM}+\sqrt{\mathcal{T}} \bgam$
with the sphere and the plate, respectively. Notice that $\Theta_\mathrm{S}$
is independent of $a$, whereas $\Theta_\mathrm{P}$ reads
\begin{align}\label{CEff-SaP-3}
\Theta_\mathrm{P}=\theta(-(a+R+z_\mathrm{CM}+\sqrt{\mathcal{T}} \bgam_{z_{\mathrm{min}}})),
\end{align}
where $\bgam_{{z_{\mathrm{min}}}}$ denotes the worldline's {extremal
  extent into the negative $z$ direction}. 

{As we are interested in calculating the Casimir force,
  $F_\mathrm{c}=-\mathrm{d}E_\mathrm{c}/\mathrm{d}a$, the derivative acting
  only on $\Theta_\mathrm{P}$ produces a $\delta$ function which eliminates the
  $z_\mathrm{CM}$ integral. The Casimir force thus simplifies to}
\begin{align}\label{CEff-SaP-4}
F_\mathrm{c}=-\frac{1}{16\pi} \,  \left\langle \int_{0}^\infty \frac{\mathrm{d}\mathcal{T}}{\mathcal{T}^{3}}
\int_0^\infty \mathrm{d}r \, r\,
\Theta_\mathrm{S} [\widetilde{\mathbf{x}}+\sqrt{\mathcal{T}} \ \widetilde{\bgam}]\right\rangle.
\end{align}
Here, we have 
introduced
\begin{align}\label{CEff-SaP-5}
\widetilde{\mathbf{x}}=\left(
                         \begin{array}{c}
                           r \\
                           0 \\
                           -a-R \\
                         \end{array}
                       \right), \ \ \ \ \ \
  \widetilde{\bgam}=\left(
                         \begin{array}{c}
                           \bgam_x \\
                           \bgam_y \\
                           \bgam_z-\bgam_{z_{\mathrm{min}}} \\
                         \end{array}
                       \right).
\end{align}
{The transition from worldline calculations of the force does not only
  lead to technical simplifications. Also, the classification of relevant
  worldlines changes slightly: for the Casimir energy in Eq. (\ref{Int-1}) the
  worldlines are scaled by the propertime $\sqrt{\mathcal{T}}$ with respect to
  their center of mass which is finally integrated over. For a given center of
  mass, all points on a worldline
  $\mathbf{x}_{\mathrm{CM}}+\sqrt{\mathcal{T}}\bgam(t_i)$ lie on rays
  originating from the center of mass. These rays are traced out by the $\pt$
  integral running from $\pt=0$ to $\pt=\infty$.}

{By contrast, the Casimir force in \Eqref{CEff-SaP-4} results from
  worldlines which are attached to the point $\widetilde{\mathbf{x}}$ on the
  plate. For a given point $\widetilde{\mathbf{x}}$, all points on a worldline
  $\widetilde{\mathbf{x}}+\sqrt{\mathcal{T}} \,
  \widetilde{\mathbf{\bgam}}(t_i)$ lie on  rays which now
  originate from $\widetilde{\mathbf{x}}$. Again these rays are traced out by
  the $\pt$ integral.} 

{Now, the plate is always touched by by construction for all values of
  $\mathcal{T}$, the remaining problem being the detection of intersection
  events with the sphere.  Adapting methods from \cite{Gies:2006cq}, it is clear that only those points of a worldline
  lying on the rays intersecting the sphere eventually pass through the sphere
  for some values of $\mathcal{T}$. Let $\{\widetilde{\bgam}(t_k)\}$ denote the
  set of points on such rays intersecting the sphere, with $k$ labeling these
  rays for a discretized worldline. Those values of propertime $\mathcal{T}$
  for which this point lies exactly on the sphere} can be obtained from the
equation
\begin{align}\label{CEff-SaP-6}
(\widetilde{\mathbf{x}}+\sqrt{\mathcal{T}}\, \widetilde{\bgam}(t_k))^2=R^2.
\end{align}
Equation \eqref{CEff-SaP-6}  has two solutions,
\begin{align}\label{CEff-SaP-7}
\mathcal{T}^{\pm}_{k}=\left(\frac{\widetilde{\mathbf{x}}\cdot \widetilde{\bgam}(t_k)}{|\widetilde{\bgam}(t_k)|^2} \mp
\sqrt{\left(\frac{\widetilde{\mathbf{x}}\cdot
    \widetilde{\bgam}(t_k)}{|\widetilde{\bgam}(t_k)|^2}\right)^2
-\frac{|\widetilde{\mathbf{x}}|^2-R^2}{|\widetilde{\bgam}(t_k)|^2}}\right)^2.
\end{align}
For $\mathcal{T}\in (\mathcal{T}^-_{k},\, \mathcal{T}^+_{k})$ the point 
$\widetilde{\mathbf{x}}+\sqrt{\mathcal{T}}\, \widetilde{\bgam}(t_k)$ lies
inside the sphere.  {The point $\widetilde{\mathbf{x}}$ can be viewed as
  a tip of a cone that wraps around the sphere with the opening angle
  $2\alpha$, with} $\sin(\alpha)=R/|\widetilde{\mathbf{x}}|$. The value of the
square root in \Eqref{CEff-SaP-7} varies between zero and $R$. The square root
is zero if the ray merely touches the sphere, and $R$ if the ray lies on the
cone's axis, i.e., if it coincides with {the direction spanned by}
$\widetilde{\mathbf{x}}$.

{For a given $r$, the worldline intersects the sphere if the propertime
  $\pt$ is in one of the intervals bounded by \Eqref{CEff-SaP-7} for all
  possible values of $k$. Denoting these intervals by
  $\overline{\mathcal{T}}_{k}:=[\mathcal{T}^-_{k},\, \mathcal{T}^+_{k}]$. The
  total support of the propertime integral then is
\begin{align}\label{CEff-SaP-8}
\mathcal{S}(r)&=\bigcup_k \overline{\mathcal{T}}_{k}.
\end{align}
The $r$ dependence of this support arises from the fact that the set of $k$ rays
lying inside the cone depends on the position $r$ where the worldline is
attached to the plate.} The Casimir force (\ref{CEff-SaP-2}) now reads
\begin{align}\label{CEff-SaP-9}
F_\mathrm{c}=-\frac{1}{16\pi} \,  \left\langle \int_0^\infty \mathrm{d}r \, r
\int_{\mathcal{S}(r)}^\infty \frac{\mathrm{d}\mathcal{T}}{\mathcal{T}^{3}} 
\
\right\rangle.
\end{align}
The most time-consuming part of the algorithm is the determination of $S(r)$
if the distance between the sphere and plate is small. To reduce the
computational time, {it is advisable to reduce the $N$ points per
  worldline to the subset of $k<N$ points on the above mentioned rays
  intersecting the sphere. For a given $r$,} all points {on rays outside
  the cone can immediately be dropped}. Furthermore in the process of taking
the $r$ integral from zero to infinity, {the opening angle of the cone
  shrinks.} All points {on rays} which leave the cone {through its
  upper half can then be dropped completely from the calculation}, as they
will never enter the cone again. {Only rays below the cone, i.e., between
  the cone and the plate, can enter the cone for larger values of $r$.} With
these optimizations and {with one integral less,} the computational time
{for Casimir force calculations} is significantly reduced compared with
{those of the Casimir energies studied in} previous worldline
investigations.

{These simplification facilitate to extend the previously studied
  parameter range to even larger $a/R$ ratios with higher statistics.}

\subsection{Cylinder above a plate}

{In many respects, the cylinder-plate configuration is ``in between'' the
  sphere-plate configuration and the classic parallel-plates case}. This also
holds for the experimental realization: the effort of keeping the cylinder
parallel to the plate is less than it is the case for two parallel plates
{\cite{BrownHayes:2005uf}; for the sphere-plate case, this issue is simply
absent. As a clear benefit,} the force can, in principle, be made arbitrarily
large, since it is proportional to the length of the cylinder.

The geometry of the cylinder-plate configuration {can be parameterized
  analogously to the preceding sphere-plate case}: we consider the symmetry
axis of a cylinder of an (infinite) length $L_y$ and radius $R$ to coincide
with the $y$ axis.  The infinite plate  lies in the $z=-(R+a)$ plane, with
$a$ being the distance between the cylinder and the plate.

The Casimir force can be obtained directly from \Eqref{Int-1}, where use the
fact that the $\Theta_\Sigma [\mathbf{x}(\tau)]$ functional factorizes
(cf. \Eqref{CEff-SaP-2}) 
\begin{align}\label{CEff-CaP-1}
\Theta_\Sigma [\mathbf{x}(\tau)] =\Theta_\mathrm{Cyl} [\mathbf{x}_\mathrm{CM}+\sqrt{\mathcal{T}} \bgam]\Theta_\mathrm{P} [\mathbf{x}_\mathrm{CM}+\sqrt{\mathcal{T}} \bgam]. 
\end{align}
Here $\Theta_\mathrm{Cyl}$ and $\Theta_\mathrm{P}$ account for the
intersection of a worldline $\mathbf{x}_\mathrm{CM}+\sqrt{\mathcal{T}}\bgam$
with the cylinder and the plate, respectively. Again, only
$\Theta_{\mathrm{P}}$ depends on $a$ and is given in \Eqref{CEff-SaP-3}.

The $y$ integral in the Casimir energy \eqref{Int-1} is now trivial due to
translational symmetry. The Casimir force can then be obtained directly from
\Eqref{CEff-SaP-4} and reads
\begin{align}\label{CEff-CaP-2}
F_\mathrm{c}=\frac{ L_y}{16\pi^2} \,  \left\langle \int_{0}^\infty \frac{\mathrm{d}\mathcal{T}}{\mathcal{T}^{3}}
\int_0^\infty \mathrm{d}r \
\Theta_\mathrm{Cyl} \left[\widetilde{\mathbf{x}}+\sqrt{\mathcal{T}} \ \widetilde{\bgam}\right]\right\rangle,
\end{align}
where {$r=|x_{\mathrm{CM}}|$ and}
\begin{align}\label{CEff-CaP-2b}
\widetilde{\mathbf{x}}=\left(
                         \begin{array}{c}
                           r \\
                           -a-R \\
                         \end{array}
                       \right), \ \ \ \ \ \
  \widetilde{\bgam}=\left(
                         \begin{array}{c}
                           \bgam_x \\
                           \bgam_z-\bgam_{z_\mathrm{zmin}} \\
                         \end{array}
                       \right).
\end{align}
As in the case of the sphere, the worldlines
$\widetilde{\mathbf{x}}+\sqrt{\mathcal{T}} \, \widetilde{\bgam}$ {are
  attached to the plate at the point} $\widetilde{\mathbf{x}}$. The only
difference is that the worldlines are now $2$ dimensional -- a fact which
reduces the computational cost.  Only those points of a worldline lying on the
rays intersecting the cylinder pass through the latter for some values of
$\mathcal{T}$. {The construction of the support of the $\pt$ integral is
  identical to that for the sphere-plate case}, such that the total Casimir
force on the cylinder can be written as in \Eqref{CEff-SaP-9}
\begin{align}\label{CEff-CaP-3}
F_\mathrm{c}=-\frac{L_y}{16\pi^2} \ \int_0^\infty \mathrm{d}r \left\langle
\int_{\mathcal{S}(r)} \frac{\mathrm{d}\mathcal{T}}{\mathcal{T}^{3}}
\right\rangle.
\end{align}

\subsection{Zero-temperature results for the Casimir force}

It is instructive to compare our results not only with analytic estimates, but
also with the much simpler proximity force approximation (PFA). The latter is
used by default for the data analysis of geometry corrections in most
experiments. It derives from a classical reasoning for generalizing the
parallel-plate case; thus, deviations of the exact result from the PFA
estimate also parameterize genuine geometry-induced quantum behavior. 

\begin{figure}[t]
\begin{center}
\includegraphics[width=0.95\linewidth]{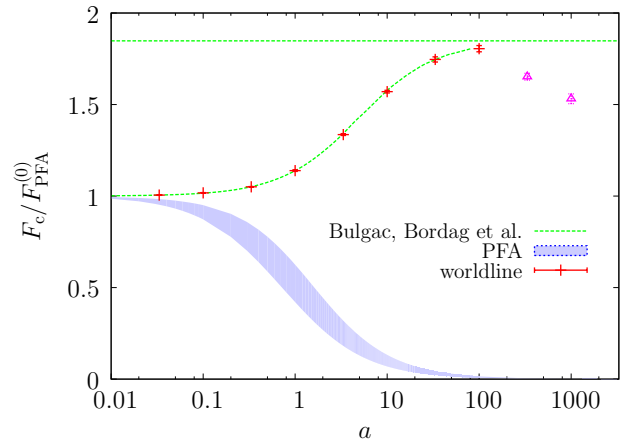}
\end{center}
\caption{Casimir Force of a sphere of radius $R=1$ above an infinite plate
  vs. the distance $a$. The force is normalized to the zeroth-order PFA
  formula. We observe an excellent agreement with the exact {asymptotic
    solutions for small $a$ \cite{Bordag:2006vc} and for large $a$
    \cite{Bulgac:2005ku} up to $a=100$.}  For larger $a$, the number of $N$
  points per loop has to be increased far beyond $N=2\cdot 10^7$ used for this
  plot; otherwise, the sphere falls through the rough mesh provided by
    the insufficiently discretized worldline, leading to a systematically
    underestimated force as is visible here for $a>100$ (pink triangles).  }\label{CEff-RT0-1}
\end{figure}

Roughly speaking, the PFA  subdivides the surfaces into small
surface elements, applies the parallel-plate force or energy law to pairs of
surface elements and integrates the resulting force density. The PFA is
inherently ambiguous as the measure for this final integration is not unique:
possible alternatives are the surface measures of one of the involved surfaces
or any intermediate auxiliary surface. {Later on, we will refer to the
  ``sphere-based'' or ``plate-based'' PFA as two generic options for the
  integration measure.} The PFA for the present configuration
is discussed in detail in Appendix \ref{app:PFA}.

The concept of the PFA can also be translated into the worldline
  picture: as an approximation to the ensemble of complicated multidimensional
  worldlines, we may reduce the worldlines to one dimensional straight lines.
The length of a line then corresponds to the average extent of a worldline
{into a certain relevant direction in a given geometry. This picture also
  explains the occurrence of deviations from the PFA as well as the sign of
  these deviations in the Dirichlet case:} due to the spatial extent of the
worldlines, they generically intersect both boundaries {for smaller
  values of $\pt$ than simple straight lines. As small propertimes yield
  quantitatively larger contributions, this property} then results in a
greater force. {More precisely}, the size squared of a worldline is
proportional to the propertime parameter which is in the denominator of the
worldline formula, see \Eqref{Int-1}.  This explains {why worldline
  results are typically underestimated by the PFA} for small separations of
the objects.  For very small separations, the upper bound of the propertime
integration can {effectively} be set to infinity, whereas the lower bound
is {a measure for} the first (proper-)time, when a worldline intersects
both objects.

\begin{figure}[t]
\begin{center}
\includegraphics[width=0.95\linewidth]{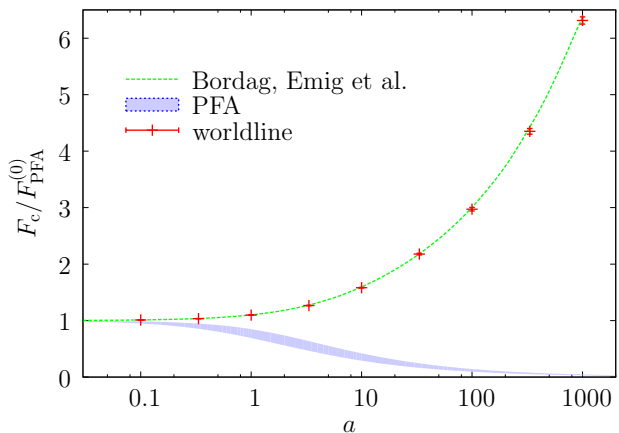}
\end{center}
\caption{Casimir Force of a cylinder of radius $R=1$ above an infinite plate
  vs. the distance $a$. The force is normalized to the zeroth-order PFA
  formula. We observe an excellent agreement with the exact {asymptotic
    solutions for small $a$ \cite{Bordag:2006vc} and for large $a$
    \cite{Emig:2006uh} up to $a=1000$.}}\label{CEff-RT0-2}
\end{figure}

The Casimir force for the sphere and cylinder is compared to the PFA
{estimates} in Fig. \ref{CEff-RT0-1} and \ref{CEff-RT0-2}
respectively. {For similar comparisons for the Casimir energy, see
  \cite{Gies:2006cq}.}  We have normalized the force to the leading-order PFA,
which {is exact in the limit of vanishing separation $a$.} The systematic
study of the PFA, also from the worldline point of view, is summarized in
Appendix \ref{app:PFA}.  We observe that the normalized force obtained with
worldline numerics does not lie inside the {range spanned by the
  ambiguity of the PFA estimates. Most prominently, the sign of the deviations
  from the $a\to0$ limit is different in the Dirichlet scalar case, as can be
  understood in the worldline picture described above. These observations have
  been frequently made in the literature before
  \cite{Gies:2003cv,Scardicchio:2004fy,Gies:2006bt,Bulgac:2005ku,Emig:2006uh}.}
 
In order to obtain Fig. \ref{CEff-RT0-1} we have used ensembles with up to
$n_\mathrm{L}=1.6\cdot 10^6$ and $N=2\cdot 10^7$. At very small distances $a$
the number of points per loop is not very important, since {part of the
  systematic error is reduced by normalizing to the leading order result;}
thus, even $N=5000$ is sufficient for example for $a=0.0333$ at a precision
level of $0.1\%$. On the other hand at $a=100$ the number of points per loop
used was $1.5\cdot 10^7$. For larger distances the number of points per loop
has to be increased far beyond $2\cdot 10^7$, since, as shown in
Fig. \ref{CEff-RT0-1}.  {Even such high resolution is not sufficient to
  resolve the small sphere for larger distances.}

{Already anticipating our results for finite temperature, this
  observation} gives us a rough estimate for the validity limits at small
temperatures. Below, we observe that for {$a/R\ll 1$} the maximum of the thermal
contribution to the force density at low temperatures {$T<1/R$}  lies outside the
sphere on scales $r\sim 1/T$. {From the fact
  that ensembles with} $N=1.5\cdot 10^7$   are reliable for those cases where
the dominant contribution to the force density lies within $r\lesssim 100$, we
conclude that temperatures above $T \gtrsim 0.01/R$
are accessible {also} in the limit $a\to 0$. 

{From an algorithmic point of view, the sphere-plate and cylinder-plate
  configurations differ with respect to computational efficiency also beyond
  the trivial dimensional factors:} for a sphere at large separations, a large
fraction of points of a worldline {can be dropped} right from the beginning, as
they never see the sphere, {i.e, they never lie on a ray inside the
  cone}. The situation is different for a cylinder. Dealing with a two
dimensional problem, we use two dimensional worldlines and the number of
points per worldline {which now have to lie in a wedge} is higher than
{those lying in a cone for the sphere-plate case}.  Using comparable
worldlines with a large number of points per loop, we thus expect the
worldline numerics to break down at far larger distances $a$ than in the case
of a sphere. This is indeed the case as is visible in Fig. \ref{CEff-RT0-2}.

For the Fig. \ref{CEff-RT0-2}  we have also used ensembles with up to
$n_\mathrm{L}=1.1\cdot 10^6$ and $N=2\cdot 10^7$. At $a=100$, the number of
points per loop used was $3\cdot 10^6$, and increased to $N=1\cdot 10^7$  for $a=333$ and
up to $N=2\cdot 10^7$ for $a=1000$. As expected, the required number of
points per loop for a certain $a$ is less here than in the case of a
sphere. Even at such large separations as $a=1000$, we observe an excellent
agreement with \cite{Emig:2006uh}.
The corresponding estimate for the validity limits at small temperatures 
then is $T>0.001/R$.

\section{Casimir effect at finite temperature}
\label{sec:fin-T}

\subsection{General considerations}

At finite temperature
$T=1/\beta$, the free energy can be decomposed into its zero-temperature part
$E_\mathrm{c}(0)$ and finite-temperature correction $\Delta E_\mathrm{c}(T)$,
\begin{align}\label{FT-1}
E_\mathrm{c}(T)=E_\mathrm{c}(0)+\Delta E_\mathrm{c}(T).
\end{align}
The same relation holds for the Casimir force
$F_\mathrm{c}(T)=
F_\mathrm{c}(0)+\Delta F_\mathrm{c}(T)$.

{Within the worldline representation of the free energy \eqref{Int-3},
the finite-temperature correction is purely driven by the worldlines with
nonzero winding number $n$. Most importantly, } the complicated
geometry-dependent part of the calculation remains the same for zero or finite
temperature.

{Let us first perform a general analysis of the thermal correction for a
  generic Casimir configuration following an argument given in
  \cite{Gies:2009nn}. We start from the assumption that the Casimir free
  energy can be expanded in terms of the dimensionless product $aT$,}
\begin{align}\label{FT-1b}
\frac{E_\mathrm{c}(T)}{E_\mathrm{c}(0)}=1+c_1 aT+c_2 (aT)^2+c_3 (aT)^3+\dots
\end{align}
No negative exponents should be present in \Eqref{FT-1b}, since the thermal
part of the energy disappears as $T\rightarrow 0$. Generically, the $T=0$
Casimir energy $E_{\mathrm{c}}(0)$ diverges for surfaces approaching contact
$a\to0$. From \Eqref{FT-1b}, we would naively expect the same for the thermal
correction. If, however, sufficiently many of the first $c_i$'s in
\Eqref{FT-1b} vanish, then the thermal part of the Casimir energy is well
behaved and without any divergence for $a\rightarrow 0$.

This turns out to be the case for two parallel plates ($c_1=c_2=0$, and
$E_{\mathrm{c}}(0) \sim 1/a^3$) and for inclined plates ($c_1=0$, and
$E_{\mathrm{c}}(0) \sim 1/a^2$) \cite{Gies:2006xe}. Consequently, an extreme
simplification arises: the low-temperature limit of the thermal correction can
be obtained by first taking the formal limit $a=0$. This was first observed in
\cite{Gies:2009nn} and then successfully applied in \cite{Bordag:2009dz}.

In the following, we argue that there is no divergence in the local thermal
force density in the limit $a\to0$ for general geometries. 
{For a generic geometry, the} $a$-divergent part can only arise from the
regions 
of contact as $a\rightarrow 0$. The divergence for these regions at $T=0$ is
due to the diverging propertime integral over $1/\mathcal{T}^{1+D/2}$ which is
bounded from below by $\sim a^2$. This is because for worldlines smaller than
$a$ the worldline functional is always
zero. 
At finite temperature the divergence in the thermal correction for
$a\rightarrow 0$ is removed since one integrates now over $\exp(-n^2 \beta^2/4
\mathcal{T})/\mathcal{T}^{1+D/2}$, which is zero for every $n>0$ in the limit
$\mathcal{T}\rightarrow 0$. The only nonanalyticity could arise from the
infinite sum. That this is not the case can {directly be verified:}
instead of integrating over the support $\mathcal{S}$, we integrate over
$\mathcal{T}$ from zero to infinity, yielding
\begin{align}\label{FT-1bb}
 \sum_{n=1}^\infty\int_0^\infty\frac{\exp\left(\frac{-n^2\beta^2}{4\mathcal{T}}\right)}{\mathcal{T}^{1+D/2}}\, \mathrm{d} \mathcal{T}=(2T)^D \Gamma(D/2) \zeta(D).
\end{align}
For {finite temperature} $T>0$, \Eqref{FT-1bb} is a finite upper bound
for the original local thermal force density. This procedure {corresponds
  to substituting the critical regions of contact} by broader (and infinitely
extended) parallel plates, see \cite{Gies:2009nn}.  The thermal contribution
is {estimated from above} by flattening the surfaces in the contact
region. The local thermal contribution to the Casimir force of the original
configuration is clearly smaller than the finite thermal contribution of
parallel plates. As the latter does not lead to divergences for $a\to 0$,
there can also be no divergence for the general curved case arising from the
contact regions. Of course, infinite geometries may still experience an
infinite thermal force, as it is the case for two infinitely extended parallel
plates, but the local thermal contribution to the force density will be
finite.  {From a practical viewpoint,} taking the limit $a\rightarrow 0$
  first simplifies the calculations considerably.

\begin{figure}[t]
\begin{center}
\includegraphics[width=0.99\linewidth]{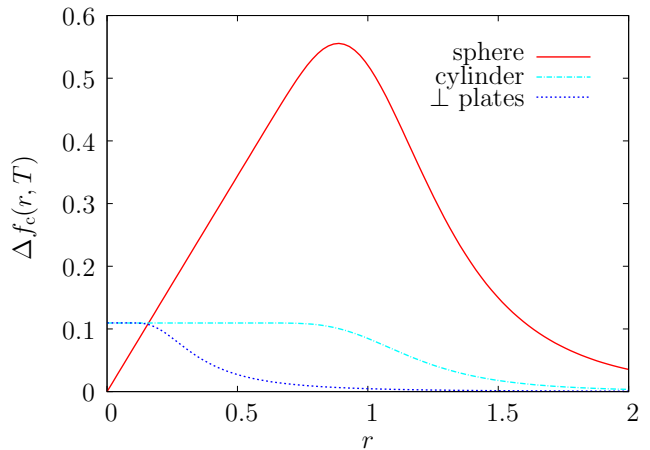}
\end{center}
\caption{The (negative) thermal force density Eq. (\ref{FT-1c}) for
  perpendicular plates (dashed blue line), {cylinder above plate (dotted
    dashed line) and sphere above plate (solid red line) both of radius $R=1$
    in the zero-distance limit $a\to 0$  for $T=1$. The sphere-plate curve
    represents the radial density including the radial measure factor $\sim
    2\pi r$.}
  The thermal force densities of cylinder and {perpendicular}
   plates at $r=0$ are
  equal to the force density of two parallel plates, $\pi^2/90\approx 0.1097$.
  The thermal force density in the sphere-plate case has a maximum of $\approx
  2\pi \times\pi^2/90$, where the factor $2\pi$ arises from the cylindrical
  measure. Note that a considerable fraction of the force density lies outside
  the sphere which only extends to $r=1$. As the temperature drops, the maximum
  moves monotonously to the right.  }\label{T0-ip-FD}
\end{figure}

Another {important} feature of low-temperature {contributions to the
  Casimir} effect is the spread of the thermal force density over regions of
size $\sim 1/T$ even for very small separations $a$. This phenomenon {has
  first been demonstrated for the configuration of two perpendicular plates at
  a distance $a$ \cite{Gies:2009nn}. (In this configuration, the sphere in
  Fig.~\ref{CEff-SaP-10} is replaced by a vertical semi-infinite plate
  extending along the positive $z$ axis and an edge at $z=0$.)  The thermal
  force density $\Delta f_\mathrm{c}(r,T)= f_\mathrm{c}(r,T)-
  f_\mathrm{c}(r,0)$ for this case as a function of the coordinate $r$ on the
  infinite surface measuring the distance from the edge (i.e., the contact
  point at $a=0$) can indeed be obtained analytically on the worldline from
  the thermal force,}
\begin{align}\label{FT-1c2}
\Delta F_\mathrm{c}(\beta)=-\frac{L_y}{16 \pi^{2}} \sum_{n=1}^\infty \left\langle \int \mathrm{d}r \!\int_{r^2/\lambda_1^2}^\infty
 \frac{e^{-\frac{n^2\beta^2}{4  \mathcal{T}}}}{\pt^{3}} \, \mathrm{d}\pt \right\rangle.
\end{align}
{Here, $\lambda_1$ is a worldline parameter measuring the extent of half
  a unit worldline, i.e., the distance measured in $x$ direction from the left
  end to the center of mass. It is clear from Fig.~\ref{CEff-SaP-10} that the
  lower bound in the $\mathcal{T}$ integral in Eq.~\eqref{FT-1c2} is given by
  $r^2/\lambda_1^2$: this is the minimal scaling value for which the worldline
  intersects the semi-infinite vertical plate}. From Eq.~(\ref{FT-1c2}), we
{read off} the following force density:

\begin{align}\label{FT-1c}
\frac{\Delta f_\mathrm{c}(r,T)}{L_y}=&-\frac{\pi^2\, T^4 }{90}\notag
\\
&+\frac{1 }{ \pi^2}\!\sum_{n=1}^\infty \! \left\langle  e^{-\frac{n^2 \lambda_1^2}{4 r^2  T^2}}\Big(\frac{ T^4 }{n^4}+\frac{ T^2\lambda_1^2 }{4  n^2 r^2}\Big)\right\rangle,
\end{align}
{Analytic results for the thermal force can be obtained by} rescaling the
radial coordinate $r\rightarrow \lambda_1 r$ per worldline and using
$\langle\lambda_1\rangle=\pi/2$. {The thermal force between the
  perpendicular plates in the limit $a\to0$ upon integration then yields}
$\Delta F_\mathrm{c}(T)=-\zeta(3)\,L_y T^3/4 \pi$ in agreement with
\cite{Gies:2008zz}.

{The perpendicular-plates configuration is special as it features a scale
  invariance in the $a\to 0$ limit: Eq.~(\ref{FT-1c}) remains invariant under
  $T\rightarrow T\alpha , \ r\rightarrow r/\alpha $ and $\Delta
  \tilde{f}_\mathrm{c}\rightarrow \Delta \tilde{f}_\mathrm{c}/\alpha^4 $ for
  arbitrary $\alpha$.  As a consequence, knowing (\ref{FT-1c}) for a single
  temperature value, say $T=1/a$, is sufficient to infer its form for all
  other $T$. Equation (\ref{FT-1c}) is shown {for $T=1/R, R=1$} in
  Fig. \ref{T0-ip-FD}.  For $r<1/T$,} the force density stays nearly constant,
corresponding to the first term in (\ref{FT-1c}). It rapidly approaches zero
for $r> 1/T$.  From this, we draw the important conclusion that the region of
constant force density in $r$ direction can be made arbitrarily large by
choosing sufficiently low $T$.

Similar consequences arise for temperature effects in other geometries.
{We plot the thermal force densities for the sphere-plate and
  cylinder-plate configuration in Fig. \ref{T0-ip-FD}.} The thermal force
density for a cylinder above a plate at $a=0$ has a shape similar to the one
of two inclined plates, whereas the radial force density of a sphere above a
plate exhibits a maximum due to the cylindric measure factor $r$, see
Fig. \ref{T0-ip-FD}. Although {these force densities are} not scale
invariant due to the additional dimensionful scale $R$ (sphere radius), its
maximum nevertheless moves away from the sphere as the temperature
drops. {We conclude that} no local approximate tools such as the PFA will
be able to predict the correct thermal force {in particular at low
  temperatures}. The fact that the force densities for sphere and cylinder are
not scale invariant leads to different temperature behaviors for $T<1/R$ and
$T>1/R$ even in the limit $a\rightarrow 0$.

\subsection{Sphere above a plate}

\begin{figure}[t]
\begin{center}
\includegraphics[width=0.99\linewidth]{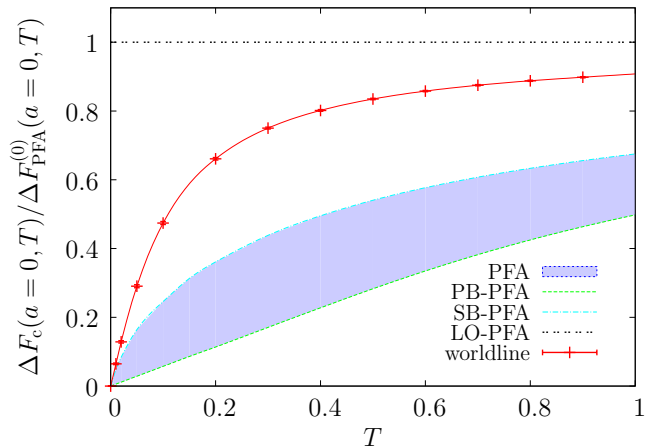}
\end{center}
\caption{Thermal Casimir force of a sphere above a plate in the limit $a\to 0$
  and for $R=1$. The worldline result and the PFA predictions are normalized
  to the leading-order PFA, cf. Eq.~(\ref{FT-SaP-3}).  The leading-order PFA
  predicts a $T^3$ behavior of the thermal force for all $T$. On the other
  hand, for small $T$ the plate-based PFA and the worldline result see a $T^4$
  behavior, whereas the sphere-based PFA sees a $T^4 \ln(T)$ one. For large
  $T\gg 1/R$ the behavior is $T^3$ for all curves.  All predictions agree in
  the large $T$ limit. For very small $T$, the worldline result runs into the
  blue area which spans the PFA predictions.
}\label{FT-SaP-F1}
\end{figure}

Let us start with the expansion of the thermal force for $a\ll R$ and for
small temperature $T\ll 1/R$. {Following our general argument given
  above, no singularities in $a$ appears in the limit $a\to 0$. Also, we
  expect that the thermal force decreases with decreasing $R$.  This motivates
  an expansion of the thermal force with only positive exponents for $a$ and
  $R$. Assuming integer exponents, dimensional analysis permits}
\begin{align}\label{FT-SaP-1}
\Delta F_\mathrm{c}(T)&=c_0 R T^3 +c_1 a T^3 \notag 
\\&+ c_2 R^2 T^4 +  c_3 a R T^4 +
\mathcal{O}\left({(a/R)^2,(TR)^5}\right). 
\end{align}
From our {numerical results in the limit $a\to0$}, we observe a $T^4$ behavior
of the thermal force, see Fig.~\ref{FT-SaP-F1}.  We conclude that $c_0\approx
0$ {is negligible} with respect to $c_2$ in the regime $T>0.01$ {where
  numerical data is available}.  {In fact, we conjecture that vanishes
  identically $c_0=0$; if so, also $c_1$ vanishes, since the configuration
  would otherwise be more sensitive to temperatures at small $a$ than at
  $a=0$. Our conjecture }is supported by the following argument {based on
}scaling properties: the {dimensionless ratio of the thermal correction and
  zero-temperature force has to be invariant under the rescaling}
\begin{align}\label{FT-SaP-1b}
a\rightarrow  a/\alpha, \ \ \  R\rightarrow R/\alpha,\ \ \ T\rightarrow  \alpha\,T.
\end{align}
The same holds for the {ratio} of $\Delta F_c (a,T)$ at $a=0$ and the
zero-temperature force at $a\neq 0$. For $a\ll R$, we can use the PFA
{for the zero-temperature force, which to leading order yields} $\sim
R/a^3$. {If} $c_0\neq 0$, {this leading ratio would be} $\sim {c_0}
(aT)^3$ which is invariant under the rescaling (\ref{FT-SaP-1b}); {in
  addition, this ratio would be invariant under (\ref{FT-SaP-1b}) with $R$
  fixed.} If $c_0=0$, then this ratio is $\sim c_2 R a^3 T^4 $ which is
invariant only under the full transformation (\ref{FT-SaP-1b}).

{The result that for $c_0\neq0$ the thermal correction would exhibit the
  same $R$ dependence as the zero-temperature force for small distances $a\ll
  R$ is counterintuitive: whereas the radial force density in the
  small-distance limit at $T=0$ is peaked right under the sphere near
  $r\simeq0$, the thermal correction arises from contributions at much larger
  $r$, cf. Fig. \ref{T0-ip-FD}. As a simple estimate, we expect that the
  thermal correction is} proportional to an effective area of the sphere,
$\approx(a+R) 2R + \pi R^2/2$, as seen by the worldlines. This estimate then
is compatible with $c_0=c_1=0$ and $c_2/c_3\approx 1.8$.

The question arises why the PFA approximation {yields} a $T^3$ behavior
despite the additional scale $R$. The reason is that $R$ appears only in the
combination $r^2/R$ in the force density, such that $T\rightarrow \alpha T$,
$r\rightarrow r/\alpha^2$ leaves the force density invariant up to a
multiplicative constant.

\begin{figure}[t]
\begin{center}
\includegraphics[width=0.99\linewidth]{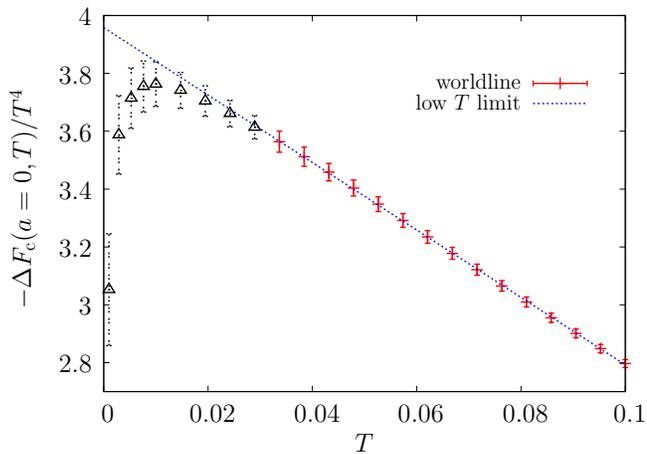}
\end{center}
\caption{Low-temperature behavior of $-\Delta F_{\mathrm{c}}(T)/T^4$ for the
  sphere-plate configuration in the limit $a\to0$ for $R=1$. For $0.03<T<0.1$,
  we observe a linear behavior which can be fitted to $\Delta
  F_{\mathrm{c}}(a=0,T)\approx -3.96\, R^2 T^4+11.66\,R^3T^5$.  We have used $40\,
  000$ loops with $2\cdot10^6$ ppl each. For $T<0.03$, the number of points per
  loop used is not sufficient to resolve the sphere properly, inducing
    systematic errors (black triangles).  }\label{FT-SaP-F2}
\end{figure}

\begin{figure}[t]
\begin{center}
\includegraphics[width=0.99\linewidth]{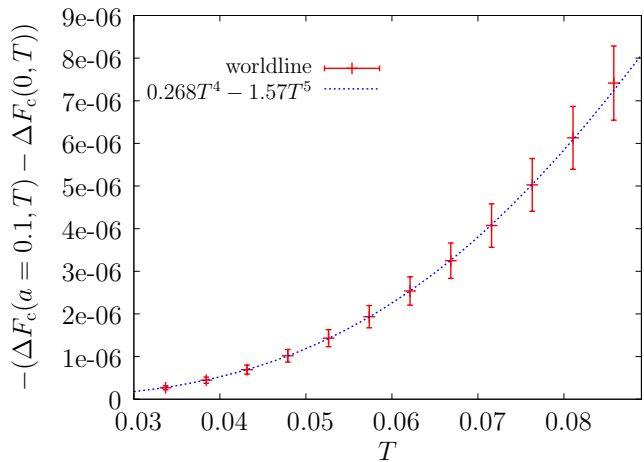}
\end{center}
\caption{
Plot of the $a$-dependent part of the (negative) thermal Casimir
  force for a sphere at $a=0.1$ and $R=1$ in the low-temperature regime. In
  the range $0.03<T<0.08$, the worldline data for $-(\Delta
  F_\mathrm{c}(0.1,T)-\Delta F_\mathrm{c}(0,T))$ is well approximated by
  $0.268\, T^4-1.57\,T^5$ {({dashed blue curve})}, {corresponding to a $c_3$
    coefficient in Eq.~(\ref{FT-SaP-1}) $c_3\approx -2.7$.} Note that the
  absolute value of the Casimir force has increased with $a$.  We have used
  $40\, 000$ loops with $2\cdot10^6$ ppl each.  }\label{FT-SaP-F3}
\end{figure}

Let us return to Eq.~(\ref{FT-SaP-1}). For $c_2$ and $c_3$, we obtain
{numerically} (see Figs.~\ref{FT-SaP-F2} and \ref{FT-SaP-F3})
\begin{align}\label{FT-SaP-2}
c_2\approx-3.96(5),  \ \ \ c_3\approx -2.7(2).
\end{align}
{These numbers can be confirmed by the exact $T$-matrix representation
  \cite{Bordag:privat}}.
Note that both coefficients have the same sign{, implying} that the
absolute value of the thermal {correction to the} Casimir force
\textit{increases} with increasing $a$ for sufficiently small $a$ {and $T$.}
This apparently anomalous behavior can be understood in geometric terms within
the worldline picture \cite{Weber:2010kc}.

\begin{figure}[t]
\begin{center}
\includegraphics[width=0.99\linewidth]{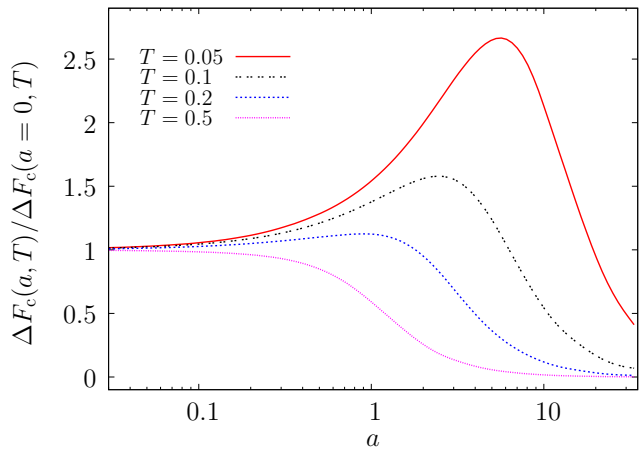}
\end{center}
\caption{Thermal correction to the Casimir force of a sphere for various
  temperatures $T$ and $R=1$, normalized to the thermal force at $a=0$. For
  sufficiently small temperatures, the absolute value of the thermal force
  correction $\Delta F(T)$ first \textit{increases} with increasing $a$. For
  $T\leq0.05$ the small $a$ behavior is well described by $1+a(2.68 R T^4-
  R^2 15.7\,T^5)/\Delta F(a=0,T)$. This verifies the fit used in
  Fig. \ref{FT-SaP-F3} with coefficient linear in $\sim a$ in front of
  $T^5$. From this prediction, we would expect the curves to be monotonically
  decreasing for $T>2.7/15.7 R \approx 0.17/R$ in the leading order.  Due to
  higher-order terms, the $T=0.2$ curve still increases slightly
  first. The statistical errors of the worldline calculation are {of} the
    order of the thickness of the curves.
}\label{FT-SaP-F4}
\end{figure}

 The system has a critical temperature $T_{\text{cr}}\simeq 0.34(1)/R$: For
 $T>T_{\text{cr}}$, the thermal force decreases monotonically for increasing
 sphere-plate separation $a$ {in accordance with} standard expectations. For
 smaller temperatures $T<T_{\text{cr}}$, the thermal force first increases for
 increasing separation, develops a maximum and then approaches zero as
 $a\rightarrow \infty$. The peak position is shifted to larger $a$ values for
 increasing thermal wavelength, i.e., decreasing temperature. In all cases,
 the force remains attractive, see Fig. \ref{FT-SaP-F4}. As an example, room
 temperature $T=300$K corresponds to the critical temperature for spheres of
 radius $R\simeq 2.6\mu$m. For larger spheres, room temperature is above the
 critical temperature such that the thermal force is monotonic. For smaller
 spheres, the thermal force is non-monotonic at room temperature. If, for
 instance, $T=70$K and $R=1.6\mu$m, the thermal force increases up to $a\simeq
 9\mu$m.

The high-temperature limit $T\gg 1/R$  agrees
with the PFA prediction for $a\to 0$ and reads
\begin{align}\label{FT-SaP-3}
\Delta F_\mathrm{c}(T\rightarrow \infty)=-\frac{\zeta(3)R}{2} T^3.
\end{align}
{In the limit $a\to 0$, the PFA yields Eq.~(\ref{FT-SaP-3})} for \textit{all}
$T$. {This is because geometrically the leading-order PFA corresponds to
  approximating the sphere by a paraboloid, which is a scale-invariant
  configuration at $a=0$.}  At finite $a$, the scale invariance is broken and
a term $\sim +R\, \pi^3 a T^4/45$ {appears on the right-hand side of}
Eq.~(\ref{FT-SaP-3}) {at low temperature in the leading-order PFA}. {By
  contrast, we observe that the true $a\to0$ limit is characterized by a $T^4$
  behavior for small $T$ and $T^3$ behavior for large $T$}. Also, {the sign of
  the correction at finite $a$ is different: the full worldline result
  predicts an increase whereas the PFA correction reduces the absolute value
  of the force}, see Fig. \ref{FT-SaP-F3}.

It is interesting to compare our results to another PFA scheme {beyond
  the leading-order PFA:} the plate based PFA. This scheme is not scale
invariant at $a=0$, as the low-temperature limit for $a\ll R$ is also
quartic and given by
\begin{align}\label{FT-SaP-4}
\Delta F^\mathrm{PFA}_{\mathrm{PB}}(a\ll R,T\rightarrow 0)=-\frac{\pi^2 T^4 }{ 90} \, \pi R^2 .
\end{align}
Equation \eqref{FT-SaP-4}, {in fact, corresponds to} the thermal force
density of two parallel plates {integrated over the area} of the region
below the sphere, $\pi R^2$. Numerically, the corresponding worldline
coefficient is more than ten times larger than the PFA prefactor $\pi^3 /90
\approx 0.345$.  In Eq.~(\ref{FT-SaP-4}), the low-$T$ behavior at finite $a$
is exponentially suppressed, implying that the plate based PFA prediction for
$c_3$ is zero -- which is again in contradiction with our worldline analysis.
The formulae (\ref{FT-SaP-3}) and (\ref{FT-SaP-4}) are derived in {the
  Appendix}.  The thermal force at $a=0$ is shown together with the PFA
predictions in Fig. \ref{FT-SaP-F1}.

\begin{figure}[t]
\begin{center}
\includegraphics[width=0.99\linewidth]{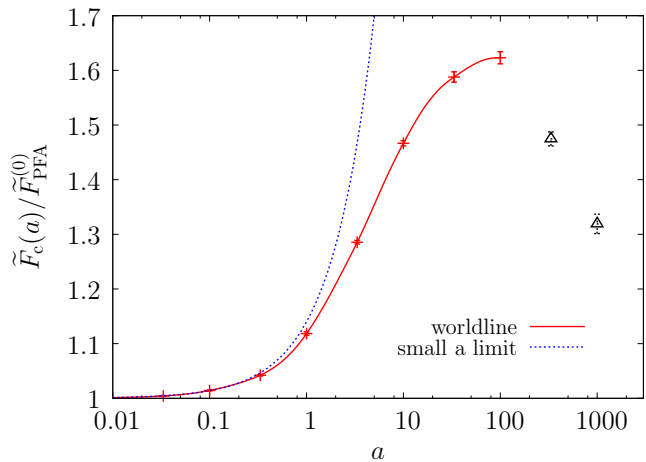}
\end{center}
\caption{High-temperature coefficient $\widetilde{F}_\mathrm{c}(a)$ for the
  sphere-plate configuration normalized to the corresponding PFA coefficient.
  At large $a$, the normalized coefficient {is conjectured to} approach a
  constant. For small $a$, the behavior is well described by $1+(0.14\pm
  0.015)\,a$.  {For $a>100$, systematic errors similar to those of
    Fig.~\ref{CEff-RT0-1} set in (black triangles); at finite temperature,
    these errors are even more pronounced due to a softer propertime exponent
    $5/2$.}
}\label{FT-SaP-F5}
\end{figure}

We now turn to the high-temperature limit, {in a strict sense
  corresponding to $T\gg 1/a$ and $T\gg 1/R$. The second requirement} is
automatically fulfilled in the {small-distance limit $a\ll
  R$}. {Quantitatively, it turns out that the high-temperature regime is
  already approached for $T\gg 1/a$ and $T\ll 1/R$.} 

A special case arises for $a\to 0$, where the high-temperature limit agrees
with the PFA prediction Eq.~(\ref{FT-SaP-3}) in the leading order. For $a> 0$,
the high-temperature limit is linear in $T$ and the total force becomes
classical, i.e., independent of $\hbar c$. 
This behavior is rather universal being a simple consequence of {\em
    dimensional reduction} in high-temperature field theories, or
  equivalently, of the linear high-temperature asymptotics of bosonic thermal
fluctuations \cite{Balian:1977qr,Feinberg:1999ys,Klich:2000qm,Nesterenko:2000cf,Bordag:2001jc,Weber:2009dp}. 
In order to find the
high-temperature limit, we perform the Poisson summation of the winding
sum. The Poisson summation for an appropriate function $f$ reads
\begin{align}\label{FT-SaP-5}
\sum_{n=-\infty}^\infty f(n/T)=\sqrt{2 \pi}\, T\sum_{k=-\infty}^\infty \hat{f}(2 \pi k T),
\end{align}
where $\hat{f}$ is the Fourier transform ({including a} $1/\sqrt{2\pi}$
prefactor) of $f$. Applying Eq.~(\ref{FT-SaP-5}) to the winding sum, we obtain
\begin{align}\label{FT-SaP-6}
&{2 \sum_{n=1}^\infty} \exp\left(-\frac{n^2\beta^2}{4\mathcal{T}}\right)\\
&=-1+  2 T \sqrt{\mathcal{T}\pi} 
+4 T \sqrt{\mathcal{T}\pi}\sum_{k=1}^\infty\exp\left(-\mathcal{T} (2 \pi k T)^2\right),\notag
\end{align}
For finite $a$, the propertime integral is bounded from below and the last
term is exponentially vanishing as $T\rightarrow \infty$. Evaluating the
worldline integrals for the first two terms, we obtain
\begin{align}\label{FT-SaP-7}
\Delta F_\mathrm{c}(a,T)=-F_\mathrm{c}(a)+ T \widetilde{F}_\mathrm{c}(a).
\end{align}
The evaluation of $\widetilde{F}(a)$ is analogous to  Eq.~(\ref{CEff-SaP-8}),
\begin{align}\label{FT-SaP-8}
\widetilde{F}_\mathrm{c}(a)=-\frac{ 1}{8\sqrt{\pi}} \,  \left\langle \int_0^\infty
\mathrm{d}r \, r \int_{\mathcal{S}(r)}^\infty
\frac{\mathrm{d}\mathcal{T}}{\mathcal{T}^{5/2}}, 
\
\right\rangle.
\end{align}
where the support $\mathcal{S}(r)$ is the same as in the $T=0$ case, see
Eqs.~(\ref{CEff-SaP-7}) and (\ref{CEff-SaP-8}).

\begin{figure}[t]
\begin{center}
\includegraphics[width=0.99\linewidth]{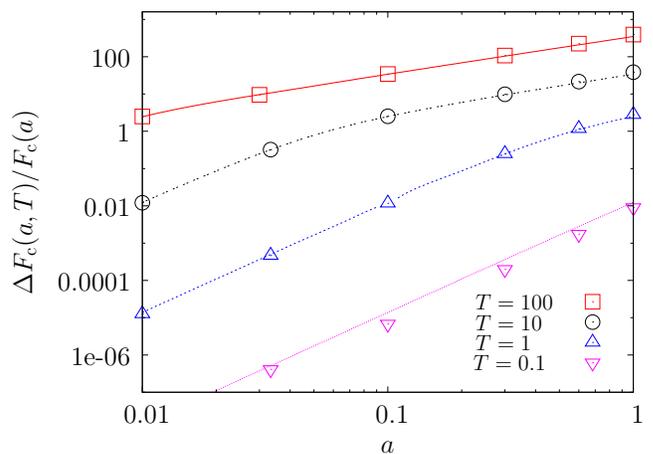}
\end{center}
\caption{{Thermal correction to the Casimir force for a sphere with $R=1$
    normalized to the zero temperature force for various temperatures and
    $a<R$. The worldline results (symbols) should be compared with the PFA
    estimate (lines). We observe that this ratio of thermal to
      zero-temperature force is
    surprisingly well described by the PFA for a wide parameter
    range, especially in the high-temperature regime. This happens because
    both $\widetilde{F}_\mathrm{c}(a)$ and $F_\mathrm{c}(a)$ increase with
    respect to the PFA with roughly the same rate, see Fig. \ref{CEff-RT0-1}
    and Fig. \ref{FT-SaP-F5}.}  }
\label{FT-SaP-F6}
\end{figure}

The Casimir force remains attractive also for high temperatures. The function
$\widetilde{F}_\mathrm{c}(a)$, normalized to the PFA prediction
\begin{align}\label{FT-SaP-9}
\widetilde F^\mathrm{PFA}_{\mathrm{c}}(a)&=-\frac{ R\, \zeta(3)   }{8 a^2},
\end{align}
is shown in Fig.~\ref{FT-SaP-F5}.  The function $a^2
\widetilde{F}_\mathrm{c}(a)$ is monotonically increasing on $0<a<100$ (similar
to $a^3 F_\mathrm{c}(a)$). At small $a$, we obtain
\begin{align}\label{FT-SaP-10}
\frac{\widetilde{F}_\mathrm{c}(a)}{\widetilde
  F^\mathrm{PFA}_{\mathrm{c}}(a)}=1+(0.14\pm 0.015)a.
\end{align}
{In analogy to the zero-temperature force, we conjecture} that also $a^2
\widetilde{F}_\mathrm{c}(a)$ remains monotonically increasing and finally
approaches a constant for $a\rightarrow \infty$. A consequence of this conjecture
is that the high-temperature limit then has a simple form, $T\gg 1/a$, without
any relation to $R$. Indeed, demanding
\begin{align}\label{FT-SaP-11}
{\Delta}F_\mathrm{c}(a,T)=-F_\mathrm{c}(a)+T \widetilde{F}_\mathrm{c}(a,T)<0
\end{align}
for a fixed $T$, { the limit $a\rightarrow \infty$ corresponds
  immediately to $T\gg 1/a$}, since $a^3 F_\mathrm{c}(a)$ itself approaches a
constant. {Our numerical data shown in Fig.~\ref{FT-SaP-F5} is indeed
  compatible with this conjecture. However, the large-$a$ limit is difficult
  to assess due to the onset of systematic errors for $a>100$.} 

Comparing Fig. \ref{CEff-RT0-1} and \ref{FT-SaP-F5}, we notice that the
  zero-temperature force $F_\mathrm{c}(a)$ and the high temperature
  coefficient $\widetilde{F}_\mathrm{c}(a)$ behave similarly. This is
  not surprising since $\widetilde{F}_\mathrm{c}(a)$ in $D=4$ Minkowski
    space corresponds to $F_\mathrm{c}(a)$ in $D=3$ Euclidean space
  due to dimensional reduction in the high-temperature limit. For finite $a$,
  the high-temperature limit is already well reached for $T\gtrsim1/2 a$.  In
  the PFA approximation, the weaker thermal force at not too small
  temperatures is normalized by the weaker zero-temperature force,
  leading to an accidental cancellation, such that for $T\gtrsim1/2a$
\begin{align}\label{FT-SaP-12}
\frac{\Delta{\widetilde{F}_\mathrm{c}}(a)}{F_\mathrm{c}(a)} \, T \approx \frac{90\,\zeta(3)}{\pi^3}\, aT\approx 3.49 \,aT,
\end{align}
independently of $R$. A comparison between the full worldline result and
  the PFA for the normalized force is shown in Fig. \ref{FT-SaP-F6} for
various $a$ and $T$.  Since for small separations $a<R$, the PFA is a
reasonable approximation already at medium temperature $1/2 a>T>1/2 R$, see
Fig. \ref{FT-SaP-F1}, we observe that the ratio between thermal
Casimir force and zero-temperature result is surprisingly well described by
the PFA for quite a wide parameter range. We stress that the PFA is
  inapplicable for each quantity alone.

\subsection{Cylinder above a plate}

{Analogous to the sphere-plate case,} we start with the expansion of the
thermal force at low temperature $T$ and {for $a\ll R$} as in
Eq.~(\ref{FT-SaP-1b}). Again, we allow only for positive exponents for $a$ and
$R$.
{Even though $\sqrt{R}$ terms appear in an  $a/R\ll1$ expansion at zero
  temperature, our numerical results at small finite temperatures, somewhat
  surprisingly, are consistent with an expansion of the type}
\begin{align}\label{FT-CaP-1}
\frac{\Delta F_\mathrm{c}(T)}{L_y}= c_2 R  T^4 +  c_3 a T^4 + \mathcal{O}\left(T^{9/2}\right).
\end{align}
The {potential leading-order terms} $c_0 \sqrt{R} T^{7/2}$ and $c_1 \sqrt{a}
T^{7/2}$ {are expected to be zero similar to the sphere-plate case, see
  above,} since the configuration of a cylinder above a plate is not invariant
unter $a\rightarrow a/\alpha$ and $T\rightarrow\alpha T$. We {have no evidence
  for } a term $\sim \sqrt{a R} T^4 $, which would lead to a nonanalytic
increase of the force. Thus at small temperatures, the powers of $T$ are found
to be integers in leading order. {Similar to the sphere-plate} case,
we expect the low-temperature contributions to the thermal force to be
proportional to the effective area $\approx L_y (2 R+a)$ seen by the distant
worldlines. {This results in the rough estimate} $c_2/c_3\approx 2$,
which also implies that both coefficients have the same sign.

\begin{figure}[t]
\begin{center}
\includegraphics[width=0.99\linewidth]{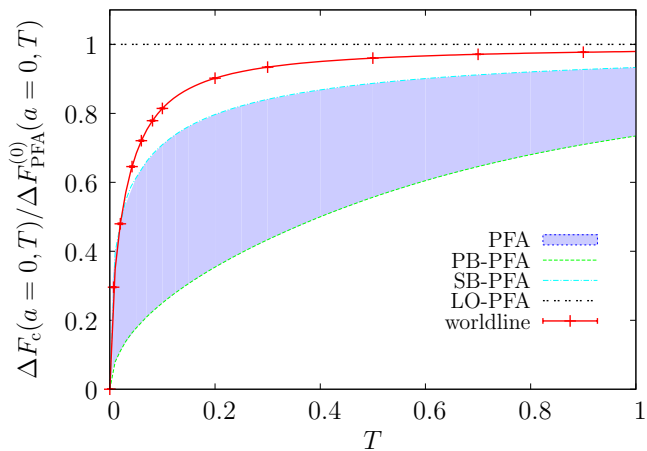}
\end{center}
\caption{Normalized thermal contribution to the Casimir force of a cylinder
  above a plate for $R=1$ in the limit $a\to 0$. The worldline result and the
  PFA predictions are normalized to the leading-order PFA
  Eq.~(\ref{FT-CaP-3}).  The leading-order PFA predicts a $T^{7/2}$ behavior
  of the thermal force for all $T$. On the other hand, {worldline
    numerical data is compatible with} a $T^4$ behavior for small $T\ll 1/R$
  and a $T^3$ behavior for large $T>1/R$. {This is also observed in the
    plate-based PFA, whereas the cylinder based PFA goes as $T^4 \ln(T)$ for
    small $T$.  All predictions agree in the high-temperature limit.} The
  worldline result enters the blue area for small $T$ into, which is the
  region spanned by the different PFA approximations.  }\label{FT-CaP-F1}
\end{figure}

\begin{figure}[t]
\begin{center}
\includegraphics[width=0.99\linewidth]{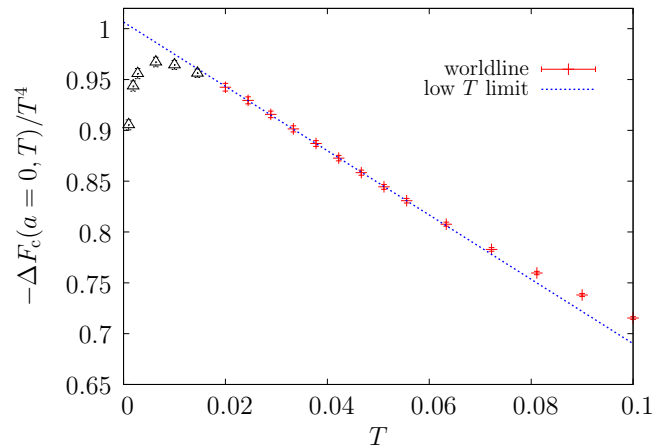}
\end{center}
\caption{Low-temperature behavior of $-\Delta F_{\mathrm{c}}(T)/T^4$ for a
  cylinder above a plate for $R=1$ in the limit $a\to 0$. For $0.02<T<0.06$,
  we observe a linear behavior. In the range $0.03<T<0.05$, our data can be
  fitted to the form $\Delta F_{\mathrm{c}}(a=0,T)/L_y\approx -1.0065\, R
T^4+3.163\,R^2T^5$.  For $T<0.02$, {systematic errors due to worldline
discretization artifacts lead to a fast decrease of the data {(black triangles)}.  We have used}
$44\, 000$ loops with $2\cdot10^6$ ppl each.  }\label{FT-CaP-F2}
\end{figure}

\begin{figure}[t]
\begin{center}
\includegraphics[width=0.99\linewidth]{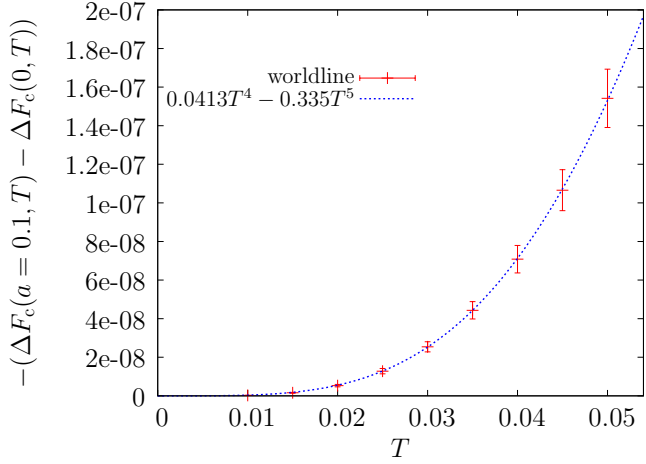}
\end{center}
\caption{Plot of the $a$-dependent part of the (negative) thermal
  {contribution to the} Casimir force for the cylinder-plate
  configuration at $a=0.1$ and $R=1$ in the low-temperature regime. In the
  range $0.025<T<0.05$, the worldline data for $-(\Delta
  F_\mathrm{c}(0.1,T)-\Delta F_\mathrm{c}(0,T))$ is well approximated by
  $0.04125\, T^4-0.335187\,T^5$.  We thus conclude that the $c_3$ coefficient
  in Eq.~({\ref{FT-CaP-1}}) is $\approx -0.4125$. Note that the absolute value of
  the Casimir force has increased with $a$.  We have used $44\, 000$ loops with
  $2\cdot10^6$ ppl each.  }\label{FT-CaP-F3}
\end{figure}

\begin{figure}[t]
\begin{center}
\includegraphics[width=0.99\linewidth]{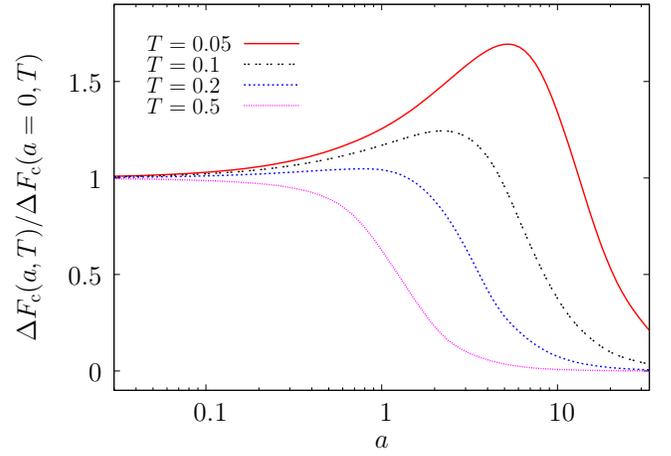}
\end{center}
\caption{Thermal {correction to the} Casimir force of a cylinder above a
  plate for various temperatures $T$ and $R=1$ normalized to the thermal
  {correction} at $a=0$. For sufficiently small temperatures, the
  absolute value of the thermal force $\Delta F(T)$ first \textit{increases}
  with increasing $a$. For $T\leq0.05$, the small-$a$ behavior is well
  described by $a(0.4125 T^4- 3.35187\,R T^5)/\Delta F(a=0,T)$. This
  verifies the fit found for Fig. \ref{FT-CaP-F3} with a coefficient linear in
  $a$ in front of $T^5$. From {this form}, we would expect the curves to
  be monotonically decreasing for $T>0.41/3.35 R \approx 0.12/R$ to
  leading order.  Due to higher-order terms the $T=0.2$ curve still
  increases slightly at the beginning.
}\label{FT-CaP-F4}
\end{figure}

{In the limit $a\to0$, our data in the regime $T>0.01$ is compatible
  with} a $T^{4}$ behavior of the thermal force. For $c_2$ and $c_3$, we
obtain (see Figs.~\ref{FT-CaP-F2} and \ref{FT-CaP-F3})
\begin{align}\label{FT-CaP-2}
{c_2\approx -1.007(7),  \ \ \ c_3\approx -0.41(4).}
\end{align}
As in the case of the sphere, both coefficients have the same sign, i.e., the
absolute value of the thermal Casimir force increases with increasing $a$ for
sufficiently small $a$ {and $T<T_\mathrm{cr}$.}  {For the critical
  temperature, we obtain $T_\mathrm{cr}\approx 0.31(1)/R$. As in the case of a
  sphere, the thermal force decreases monotonically with
  increasing $a$ for $T>T_\mathrm{cr}$; below the critical temperature, the thermal force
  first increases up to a maximum and then decreases again approaching zero
  for $a\rightarrow \infty$. The position of the maximum depends on $T$ and
  increases with inverse temperature, see Fig. \ref{FT-CaP-F4}. In both cases,
  however, the thermal force remains attractive. }

The high-temperature limit $T\gg 1/R$ agrees with the PFA 
prediction in the limit $a\to 0$ as expected,
\begin{align}\label{FT-CaP-3}
\frac{\Delta F_\mathrm{c}(T\rightarrow \infty)}{L_y}=\frac{3\, \zeta(1/2)
  \zeta(7/2)\sqrt{R}}{4\sqrt{2}\pi} T^{\frac{7}{2}} \approx -0.278 \sqrt{R} T^{\frac{7}{2}}.
\end{align}
As for the sphere, the PFA predicts the same force law (\ref{FT-CaP-3}) in the
limit $a\to 0$ for \textit{all} $T$. At finite $a$, the scale invariance is
broken and a term $\sim +0.185 a \sqrt{R} T^{9/2}$ appears on the right-hand
side of Eq.~(\ref{FT-CaP-3}) in leading-order PFA at low temperature. {By
  contrast}, we observe different power laws for different temperatures in the
limit $a\to 0$: a $T^4$ behavior for small $T$ and $T^{7/2}$ behavior for
large $T$. {Also, the sign of the finite-$a$ correction of the full result is
opposite to that of the PFA, see Fig. \ref{FT-CaP-F3}, all of which is
reminiscent to the sphere-plate case.}

Incidentally, the {beyond-leading-order} PFA schemes reflect the correct
behavior much better. We observe that the cylinder-based PFA turns out to be
the better approximation {(as for the sphere-based PFA in the preceding
  section)}. This is {opposite} to the zero-temperature case. For the
plate-based and {cylinder}-based PFA, we obtain
\begin{align}
\frac{\Delta F^\mathrm{PFA}_{\mathrm{PB}}(0,T\rightarrow 0)}{L_y}&=\frac{\pi^2
  T^4}{90} 2 R\approx -0.219\, R T^4, \label{FT-CaP-4} 
\\
\frac{\Delta F^\mathrm{PFA}_{\mathrm{CB}}(0,T\rightarrow 0)}{L_y}&=\frac{R T^4
  \left(3 \pi ^4+2 \pi ^4 \ln\left(\frac{R T}{2 \pi }\right)+180
  \zeta'(4)\right)}{90 \pi ^2}\notag
\\
&\approx
 (0.22\ln(RT/2\pi)+0.32) RT^4. \label{FT-CaP-4b}
\end{align}
The plate-based result is equal to the thermal force of two parallel plates
{integrated over an area $2 R L_y$}. The plate-based coefficient is more
than four times smaller than the worldline coefficient, whereas the leading
coefficient of the cylinder-based formula becomes arbitrarily large as $T
\rightarrow 0$.  The formulae (\ref{FT-CaP-4}) and (\ref{FT-CaP-4b}) are
derived in the appendix.  The thermal {contribution to the } force in the
limit $a\to0$ is shown together with the PFA predictions in
Fig. \ref{FT-CaP-F1}.

Let us now investigate the high-temperature limit, which can be obtained by
Poisson summation of the winding-number sum as in Eq.~(\ref{FT-SaP-6}).  A special
case arises in the limit $a\to0$, where the high-temperature limit corresponds
to the PFA prediction Eq.~(\ref{FT-CaP-3}) in leading order. For $a>0$, the
high-temperature limit is again linear in $T$ and the total force
``classical'', {i.e., independent of $(\hbar c)$,}
\begin{align}\label{FT-CaP-7}
\Delta F_\mathrm{c}(a,T)=-F_\mathrm{c}(a)+ T \widetilde{F}_\mathrm{c}(a)
\end{align}
as in Eq.~(\ref{FT-SaP-7}), we obtain
\begin{align}\label{FT-CaP-8}
\widetilde{F}_\mathrm{c}(a)=-\frac{ L_y}{8\pi^{3/2}} \,  \left\langle \int_0^\infty
\mathrm{d}r  \int_{\mathcal{S}(r)}^\infty
\frac{\mathrm{d}\mathcal{T}}{\mathcal{T}^{5/2}} 
\right
\rangle,
\end{align}
where the support $\mathcal{S}(r)$ is the same as in the $T=0$ case, see
Eq.~(\ref{CEff-CaP-3}).

\begin{figure}[t]
\begin{center}
\includegraphics[width=0.99\linewidth]{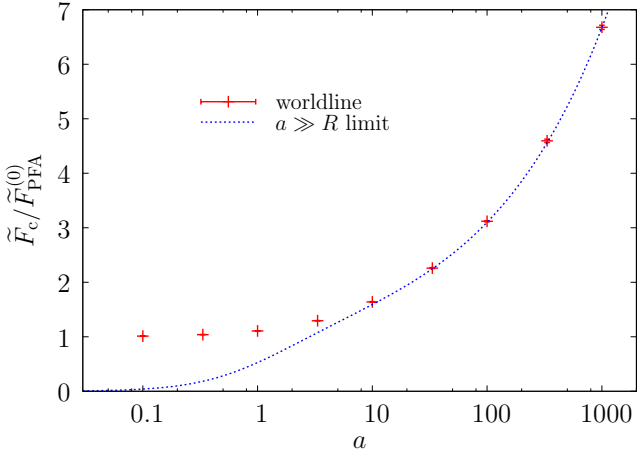}
\end{center}
\caption{High-temperature coefficient $\widetilde{F}_\mathrm{c}(a)$
  normalized to the corresponding PFA coefficient for a cylinder above a
  plate.  The large-$a$ behavior is well described by Eq.~(\ref{FT-CaP-11}). At
  small $a$, we find  a behavior $\sim 1+(0.125\pm 0.017)a$.   
}\label{FT-CaP-F5}
\end{figure}

The Casimir force remains attractive also for high temperatures. The function
$\widetilde{F}_\mathrm{c}(a)$, normalized to the PFA prediction
\begin{align}\label{FT-CaP-9}
\widetilde F^\mathrm{PFA}_{\mathrm{c}}(a)&=-\frac{ 3\sqrt{R}\, \zeta(3)   }{32 \sqrt{2} a^{5/2}},
\end{align}
is shown in Fig. \ref{FT-CaP-F5}.  The function $a^{5/2}
\widetilde{F}_\mathrm{c}(a)$ is monotonically increasing for $0<a<1000$ and is
{reminiscent} to $a^{7/2} F_\mathrm{c}(a)$. At small $a$, we obtain
\begin{align}\label{FT-CaP-10}
\frac{\widetilde{F}_\mathrm{c}(a)}{\widetilde
  F^\mathrm{PFA}_{\mathrm{c}}(a)}=1+(0.125\pm 0.017)a. 
\end{align}
At large $a$, we find using Eq.~(\ref{FT-SaP-11}) and the {analytical}
zero-temperature law \cite{Emig:2006uh},
\begin{align}\label{FT-CaP-11}
\frac{\widetilde{F}_\mathrm{c}(a)}{\widetilde
  F^\mathrm{PFA}_{\mathrm{c}}(a)}\simeq1.46(2) \frac{a^{5/2}}{(a+R)^2 \ln(a+R)}. 
\end{align}

\begin{figure}[t]
\begin{center}
\includegraphics[width=0.99\linewidth]{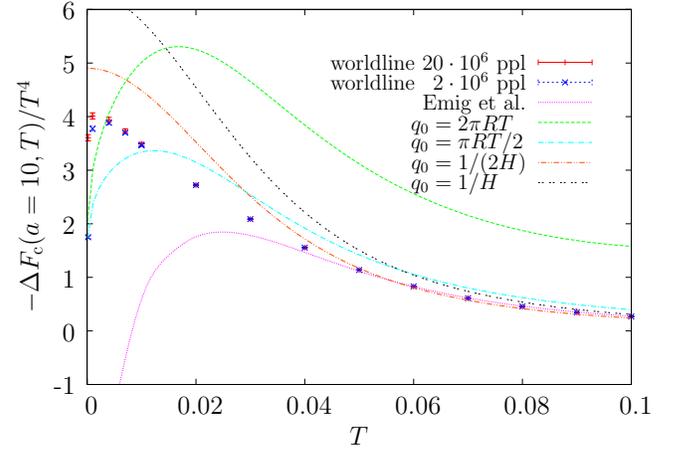}
\end{center}
\caption{Thermal contribution to the Casimir force for a cylinder above a
  plate for $a=10$ and $R=1$ { compared with the analytic result
    \eqref{FT-CaP-12b} (``Emig et al.'' \cite{Emig:2006uh}) and {various approximations as
      discussed in the text. Here, we use the abbreviation
      $H=R+a$}. Remarkably, {our proposed estimates using $q_0=\pi R T/2$
      and $q_0=1/2 H$}, cf.  Eq. (\ref{FT-CaP-12c}), describe the actual
    behavior far better than the analytic result (\ref{FT-CaP-12b}), which
    changes sign as $T\rightarrow 0$.  Also the $T>1/(R+a)$ approximation
    using $q_0=1/2H$ remains {a reasonable estimate even for $T<1/(R+a)$.
      For the worldline data,} we have used $5000$ loops with $2\cdot 10^7$
    ppl and $7000$ loops with $2\cdot10^6$ ppl.}}\label{FT-CaP-F6}
\end{figure}

\begin{figure}[t]
\begin{center}
\includegraphics[width=0.99\linewidth]{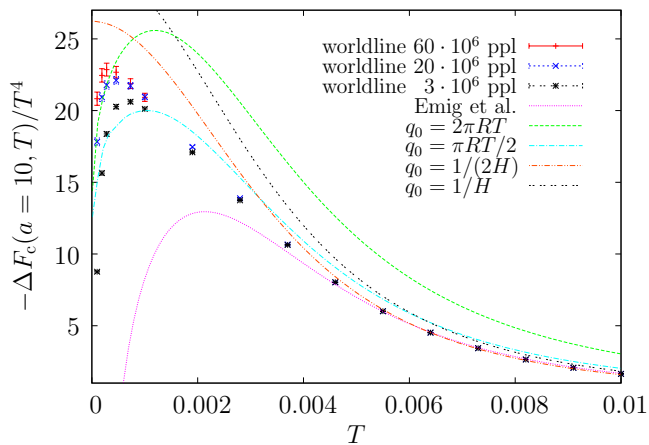}
\end{center}
\caption{Thermal {contribution to the Casimir force for a cylinder above a
    plate for $a=100$ and $R=1$ compared with the analytic result
    \eqref{FT-CaP-12b} (``Emig et al.'' \cite{Emig:2006uh}) and {various approximations as
      discussed in the text}.  Even at such large separations, we observe that
    the analytic result (\ref{FT-CaP-12b}) becomes invalid {and even
      changes sign} as the temperature approaches zero. Incidentally our
    {simple} $T>1/(R+a)$ approximation {using} $q_0=1/2 H$ describes
    the actual behavior {rather well also for smaller temperatures. For
    the worldline data,} we
    have used $2500$ loops with $6\cdot 10^7$, $5000$ loops with $3\cdot 10^7$
    ppl and $14000$ loops with $3\cdot10^6$ ppl. } }\label{FT-CaP-F7}
\end{figure}

{We can compare our results with those of an analytical result
  \cite{Emig:2006uh} in the limit $R\ll H=R+a$. The leading-order thermal
  contribution to the Casimir force in this computation based on scattering
  theory reads}
\begin{align}\label{FT-CaP-12b}
\Delta F_\mathrm{c}(a,T)=L_y  T\int_0^\infty \frac{ qe^{-2 q (R+a)}\ \text{st}(q)}{\ln(q R)}\,\mathrm{d}q,
\end{align}
{where the integrand has been approximated to leading order in $\ln^{-1}(qR)$.
Here,} $\text{st}(q/T)$ is a $2\pi T$ periodic sawtooth function which in the
range from $0$ to $2\pi T$ is given by $-q/(2\pi T)+1/2$.  {The authors of
\cite{Emig:2006uh} have given a simple estimate of the integral for the limit
$R\ll 1/2\pi T$ by replacing $\ln(q R)$ by $\ln(2\pi
R T)$ and carrying out the resulting integral. }
{We compare our worldline results with \Eqref{FT-CaP-12b} as well as with 
the simple estimate} in Fig. \ref{FT-CaP-F6} and \ref{FT-CaP-F7}.

Here, we propose {another} estimate {which is valid} for arbitrary
$T>1/(R+a)$. {In this case,} the sawtooth function is approximately
constant for $q<1/(R+a)$. We approximate the logarithm by {inserting} the
value $q_0$ for which $q\exp(-2q(R+a))$ is maximal: $q_0=1/2(R+a)$.  In turn
for $T<1/(R+a)$, the logarithm can be approximated by {insertion of the
  value} $q_0$ where $q\exp(-2q(R+a)) \text{st}(q)$ has its first maximum:
$q_0=\pi T/2$. We choose the first maximum, as the integrand is oscillating
for $q>q_0${, such that cancellation can be expected }to occur.  However,
choosing $q_0\sim T$ always leads to a regular $T^4/\ln(T)$ behavior for small
$T$, whereas Eq. (\ref{FT-CaP-12b}) changes sign at very small $T$, see
Figs.~\ref{FT-CaP-F6} and \ref{FT-CaP-F7}. We thus conclude that
Eq.~(\ref{FT-CaP-12b}) is valid for not too small  $T$.
The thermal contribution to the Casimir force then reads
\begin{align}\label{FT-CaP-12c}
\Delta F_\mathrm{c}(a,T)=\frac{-T L_y}{\ln(q_0)}\frac{\mathrm{d}}{\mathrm{d}a}\frac{ \text{coth}(2\pi(R+a) T)-\frac{1}{2 \pi (R+a)  T}}{8  (R+a)},
\end{align}
where $q_0=2\pi R T$ in the Emig et al. approximation \cite{Emig:2006uh},
whereas $q_0=R/2(a+R)$ for $T>1/(R+a)$ and $q_0=R\pi T/2$ for $T<1/(R+a)$ in
the approximation proposed here. See Figs.~\ref{FT-CaP-F6} and \ref{FT-CaP-F7}
for the results at $a=10 R$ and $a=100 R$ respectively.  In the small $T$
limit, Eq.~(\ref{FT-CaP-12c}) reads
\begin{align}\label{FT-CaP-13}
\Delta F_\mathrm{c}(a,T)=L_y\frac{2 \pi ^3 (a+R) T^4}{45 \text{ln}(q_0)}.
\end{align}
{Writing} this as $c(T,a,R)T^4$, the $T^4$ coefficient $c$ always
disappears for $q_0\sim T$ as $T\rightarrow 0$. {In our numerical
  worldline analysis, the systematic discretization errors lead to a vanishing
  of the corresponding coefficient as well, since the number of points per
  worldline becomes insufficient for resolution of the cylinder at very small
  $T$. For an increasing number of points per worldline, however, our data
  actually appears to point to a non-vanishing coefficient, see
  Figs.~\ref{FT-CaP-F6} and \ref{FT-CaP-F7}. In any case, we expect the
  leading-order multipole expansion which is behind the asymptotic result
  \eqref{FT-CaP-12b} to break down at low temperatures due to the geothermal
  interplay. }

\begin{figure}[t]
\begin{center}
\includegraphics[width=0.99\linewidth]{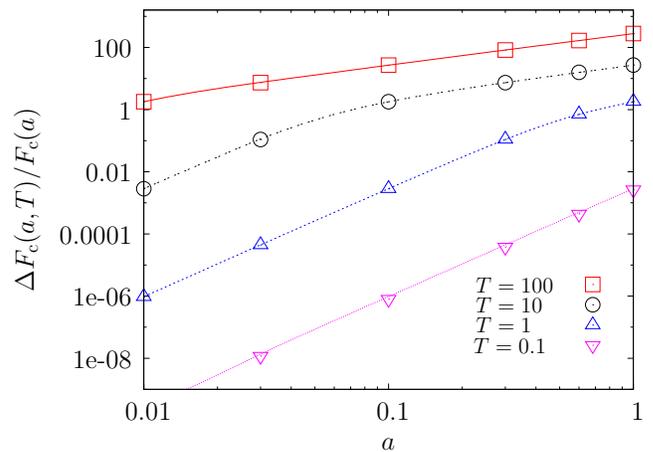}
\end{center}
\caption{{Thermal correction to the Casimir force for a cylinder with
    $R=1$ normalized to the zero-temperature force for various temperatures
    and $a<R$. The worldline results (symbols) are in a better agreement with
    the PFA estimates (lines) than in the case of a sphere, see
    Fig. \ref{FT-SaP-F6}. The normalized PFA results agree with the worldline
    results for not too low $T$ because again both
    $\widetilde{F}_\mathrm{c}(a)$ and $F_\mathrm{c}(a)$ increase with respect
    to the PFA with roughly the same rate, see Fig. \ref{CEff-RT0-2} and
    Fig. \ref{FT-CaP-F5}.}  }
\label{FT-CaP-F8}
\end{figure}

{For large $T$ on the other hand, Eq.~(\ref{FT-CaP-12c}) becomes
\begin{align}\label{FT-CaP-14}
\Delta F_\mathrm{c}(a,T)=-\frac{L_y \left(1-T\pi(a+R)\right)}{8 \pi \ln(q_0) (R+a)^3}.
\end{align}
{We observe} that the negative of the $T$-independent part
{approaches} the zero-temperature limit of the Casimir force for large
$a/R$ faster {if we choose $q_0=1/(R+a)$ rather than $q_0=1/2(R+a)$.
This choice of $q_0=1/(R+a)$ then constitutes our second estimate for
$T>1/(R+a)$.}}

{

For not too small $T$, the analytic result and the various $q_0$
approximations nicely agree with our worldline data, see Figs.~\ref{FT-CaP-F6}
and \ref{FT-CaP-F7}.  For higher temperature, the behavior becomes $\sim T$
and the different results {acquire} different slopes which {partly
  disagree} for $T\rightarrow \infty$. For $a=10$, the analytic result becomes
$\sim 0.000424 L_y T$, the $q_0=1/(R+a)$ approximation yields $\sim 0.000431
L_y T$, and the $q_0=1/2(R+a)$ approximation $\sim 0.000334 L_y T$. The
numerical worldline result is $\sim 0.00041(3) L_y T$. }

{ At $a=100$ and high $T$, the analytic result is $\sim 2.57\cdot 10^{-6} L_y
  T$, the $q_0=1/(R+a)$ approximation yields $\sim 2.66 \cdot 10^{-6} L_y T$ and the
  $q_0=1/2(R+a)$ approximation $\sim 2.31 \cdot 10^{-6} L_y T$.  The worldline
  result is $\sim 2.4(9) \cdot 10^{-6} L_y T$.

For large $a+R$, the temperature coefficient becomes $0.125 /(R+a)^2\ln(R+a)$
for both $q_0=1/(R+a)$ and $1/2(R+a)$. For the analytic result the
corresponding prefactor is greater than $0.123$ and may become $0.125$ for
$H\rightarrow\infty$. The corresponding worldline prefactor is
$0.116(2)$, see
Eq. (\ref{FT-CaP-11})}.

Let us return to the high-temperature discussion, and finally remark that also
in the case of a cylinder the thermal Casimir force normalized to the zero
temperature result is well described by the PFA for
$T\gtrsim1/2a$. Analogously to Eq.~(\ref{FT-SaP-12}), we conclude from the
dimensional-reduction argument, that the ratio of thermal to
  zero-temperature force in the high-temperature limit $T\gtrsim1/2a$ is
approximately
\begin{align}\label{FT-CaP-15}
  \frac{\Delta{\widetilde{F}_\mathrm{c}}(a)}{F_\mathrm{c}(a)} \, T \approx
  \frac{72\,\zeta(3)}{\pi^3}\, aT\approx 2.79 \,aT.
\end{align}
Also at medium temperatures this ratio is surprisingly well
described by the PFA, even better than in the case of a sphere, see Fig.
\ref{FT-SaP-F6}.  The normalized thermal force is shown in
Fig. \ref{FT-CaP-F8} for various $a$ and $T$.

\section{Conclusions}
\label{sec:conclusions}

In this work, we have analyzed the geometry-temperature interplay in the
Casimir effect for the case of a sphere or a cylinder above a plate. Since
finite-temperature contributions to the Casimir effect are induced by a
thermal population of the fluctuation modes, the geometry has a decisive
influence on the thermal corrections as the mode spectrum follows directly
from the geometry. A strong geometry-temperature interplay can generically be
expected whenever the length scale set by the thermal wavelength is comparable
to typical geometry scales. 

Within our comprehensive study of the Casimir effect induced by Dirichlet
scalar fluctuations for the sphere-plate and cylinder-plate geometry, we
observe several signatures of this geometry-temperature interplay: the thermal
force density is delocalized at low temperatures. This is natural as only
low-lying long-wavelength modes in the spectrum can be thermally excited at
low $T$. As a consequence, the force density is spread over length scales set not
only by the geometry scales but also by the thermal wavelength. This implies
that local approximation techniques such as the PFA are generically
inapplicable at low temperatures. Quantitatively, the low-temperature force
follows a $T^4$ power law whereas the leading-order PFA correction predicts a
$T^3$ behavior -- a result which has often been used in the analysis of
experimental data. Only for ratios of thermal to zero-temperature forces, we
observe a potentially accidental agreement with the PFA prediction for larger
temperatures. Here, the errors introduced by the PFA for the aspect of
geometry appear to cancel, whereas the thermal aspects might be included
sufficiently accurately. 

Another signature of this geometry-temperature interplay is the occurrence of
a non-monotonic behavior of the thermal contribution to the Casimir
force. Below a critical temperature, this thermal force first grows for
increasing distance and then approaches zero only for larger distances. This
phenomenon is not related to a competition of polarization modes as in
\cite{Rodriguez:2007,Rahi}, but exists already for the Dirichlet scalar
case. The phenomenon can be understood within the worldline picture of the
Casimir effect \cite{Weber:2010kc} being triggered by a reweighting of
relevant fluctuations on the scale of the thermal wavelength. From this
picture, it is clear that the phenomenon is not restricted to spheres or
cylinders above a plate; we expect it to occur for general compact or
semi-compact objects in front of surfaces, as long as the lateral surface
extension is significantly larger than the thermal wavelength. In fact,
another consequence of the delocalized force density is that edge effects due
to finite plates or surfaces will be larger for the thermal part than for the
zero-temperature force.

Our results have been derived for the case of a fluctuating scalar field
obeying Dirichlet boundary conditions on the surfaces. For different fields or
boundary conditions, the temperature dependence can significantly differ from
the quantitative results found in this work. This is only natural as different
boundary conditions can strongly modify the fluctuation spectrum. For
instance, the thermal part of the free energy in the sphere-plate case
exhibits different power laws for Dirichlet or Neumann boundary conditions in
the low-temperature and small-distance limit \cite{Bordag:2009dz}.  For future
realistic studies of thermal corrections, all aspects of geometry,
temperature, material properties, boundary conditions and edge effects will
have to be taken into account simultaneously, as their mutual interplay
inhibits a naive factorization of these phenomena.

\appendix

\section{The proximity-force approximation (PFA)}
\label{app:PFA}

\begin{figure}[t]
\begin{center}
\includegraphics[width=0.99\linewidth]{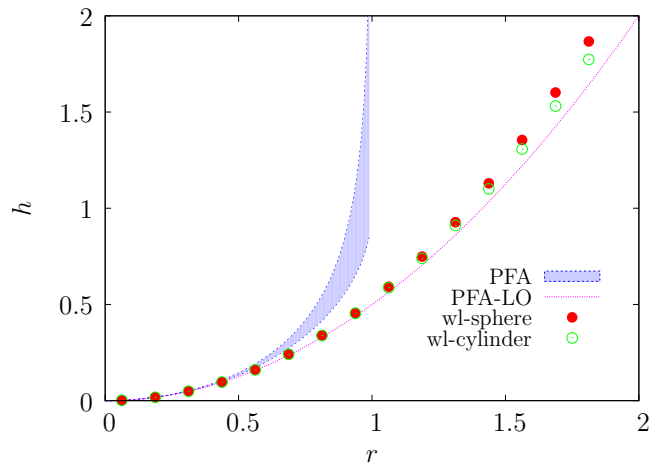}
\end{center}
\caption{The effective heights predicted by the PFA for $a=0$, $R=1$ and $T=0$
  compared with those of worldline numerics for a sphere and a cylinder,
  respectively, obtained from the $T=0$ force density. The PFA predictions lie
  in the blue area which is bounded by Eq. (\ref{A-PFA-3a}) from below and by
  Eq. (\ref{A-PFA-3e}) from above. The effective heights seen by worldlines are
  well approximated by the leading order PFA for $r$ not too large. We
  conclude that for small $a/R$ the force is described best by the leading
  order order PFA, since for small $a/R$ the force density is concentrated
  around $r=0$.}\label{A-F1}
\end{figure}
\begin{figure}[t]
\begin{center}
\includegraphics[width=0.99\linewidth]{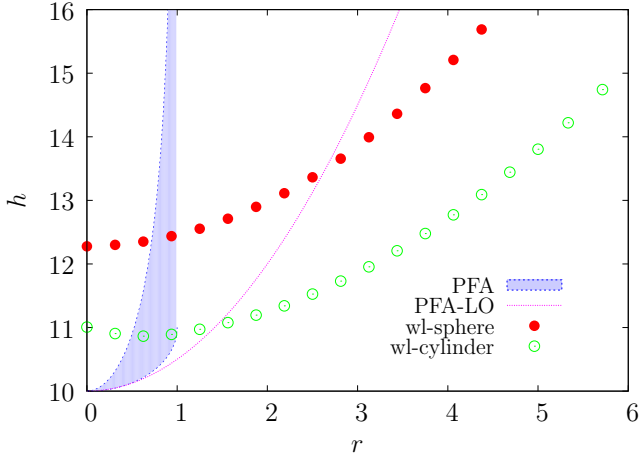}
\end{center}
\caption{The effective heights predicted by the PFA for $a=10$, $R=1$ and
  $T=0$ compared with those of worldline numerics for a sphere and a cylinder,
  respectively, obtained from the $T=0$ force density. {Note that the
    effective height for the cylinder is in a local maximum at $r=0$. }At
  greater separations the heights seen by worldlines are on average lower than
  the PFA predictions resulting in greater Casimir force. Also at larger
  separations the leading order PFA reflects best the actual situation, while
  the 'sphere/cylinder based' PFA turns out to be the worst. }\label{A-F2}
\end{figure}

The proximity force approximation is a scheme for estimating Casimir
energies between two objects. In this approach, the surfaces of the bodies are
treated as a superposition of infinitesimal parallel plates, and the Casimir
energy is approximated by
\begin{align}\label{A-PFA-1}
 E^\mathrm{PFA}(a)=\int_\Sigma \varepsilon^\mathrm{PP}(h) \, \mathrm{d}\sigma.
\end{align}
Here, one integrates over an auxiliary surface $\Sigma$, which should be chosen
appropriately. The quantity $\varepsilon^\mathrm{PP}(h)$ denotes the
energy per unit area of
two parallel plates at a distance $h$ apart, which at zero temperature reads
\begin{align}
\varepsilon^\mathrm{PP}(h)=-\frac{c_\mathrm{PP}}{h^3}\label{A-PFA-2a},
\end{align}
where $c_\mathrm{PP}=\pi^2/1440$ for the Dirichlet scalar case. 

As the PFA does not make any reference to boundary conditions, all the formulas in this appendix are analogously valid for the electromagnetic case; all force formulas then have to be multiplied by a factor of two for the two polarization modes.

At finite temperature, the corresponding expression is
\begin{align}
\frac{\Delta \varepsilon^\mathrm{PP}(h)}{c_\mathrm{PP}}&=\label{A-PFA-2b}
\\
&\!\!\!\!\!\!\!\!\!\!\!\!\!\!\!\!\!\!\!\!\!\!\!\frac{1}{h^3}-\frac{90  T}{h^2}\sum_{n=1}^{\infty}\frac{\coth(2 n \pi  h T)+2 n \pi  h T \text{csch}^2(2n \pi  h T)}{\pi^3 n^3}. \notag
\end{align}
The distance is conventionally measured along the normal to
$\Sigma$. The two extreme cases in which $\Sigma$ coincides with one of the
two bodies provides us with a region spanning the inherently ambiguous
  estimates of the PFA. 

For a sphere at a distance $a$ above a plate, we thus integrate either over the
plate ('plate based' PFA), or over the sphere ('sphere based' PFA). The
Casimir force is then obtained by taking the derivative of (\ref{A-PFA-1})
with respect to $a$. However, for the 'sphere based' PFA, $\mathrm{d}
h/\mathrm{d}a\neq 1$ (see below).
This implies that deriving the force estimate from the PFA of the energy
  in general is not the same as setting up the PFA directly for the force (the
  latter would correspond to 
  a surface integral over the parallel-plates force per unit area). In this
  work, we use the derivation via the energy (\ref{A-PFA-1}). 

The dependence of the PFA prediction on the choice of $\Sigma$
  disappears in the limit 
$a\rightarrow 0$ at zero
temperature {to leading order}. This 
result shall be called 'leading-order' PFA. It
can also be obtained by expanding the surface of the sphere/cylinder to second
order from the point of minimal distance to the plate and then using the
'plate based' PFA for this expansion.

The corresponding expressions for $h$ read
\begin{align}
h_\mathrm{PB}&=a+R-\sqrt{R^2-r^2},\label{A-PFA-3a}
\\
h_\mathrm{SB}=h_\mathrm{CB}&=\frac{a+R}{\cos(\theta)}-R,\label{A-PFA-3b}
\\
h_\mathrm{LO}&=a+\frac{r^2}{2R}.\label{A-PFA-3c}
\end{align}
For $h_\mathrm{PB}$, we integrate over $[-R,R]$, for $h_\mathrm{LO}$ over all
$r$ and for $h_\mathrm{SB}$, $h_\mathrm{CB}$ over $[-\pi/2,\pi/2]$ with an
appropriate measure. Note that right underneath the sphere/cylinder all $h$
are equal to $a$. Demanding $\mathrm{d}\theta=\mathrm{d}r$ for
$\theta\rightarrow 0$ we can transform the integration over $\theta$ into 
an integration over $r$ in a simple way by the substituting $\sin(\theta)\rightarrow
r/R$. The integral then goes from $-R$ to $R$, and the corresponding $h$ reads
\begin{align}
h_\mathrm{SB}=h_\mathrm{CB}&=-R+R\frac{a+R}{\sqrt{R^2-r^2}}.\label{A-PFA-3d}
\end{align}
Also a measure factor resulting from $R\mathrm{d}\theta=R/\sqrt{R^2-r^2}$ and
$\mathrm{d}h/\mathrm{d}a$ have to be taken into account. At zero temperature,
we can absorb these factors into the new effective height
\begin{align}
h_\mathrm{SB-eff}=h_\mathrm{CB-eff}&=\frac{\sqrt{R} \left(a+R-\sqrt{R^2-r^2}\right)}{\left(R^2-r^2\right)^{1/4}}.\label{A-PFA-3e}
\end{align}
With or without the prefactors $h_\mathrm{SB/CB}$ is always greater than
$h_\mathrm{PB}$ and $h_\mathrm{LO}$ and diverges for $r\rightarrow
R$. Since the factor approaches $1$ for small $r$, all functions $h$ coincide
in this limit.

The PFA can also be developed within worldline formalism. Calculating the Casimir
force density for two parallel plates, we have to determine that value
  of 
propertimes $\mathcal{T}$ for which one dimensional worldlines, attached to one of the
plates, touch the other plate for the first time. This event is encoded in the
lower bound of the proper time integral, whereas the upper bound is set to
infinity. Thus, we obtain
\begin{align}\label{A-PFA-4}
f_\mathrm{c}^\mathrm{PP}(h,\beta)=-\frac{1}{32 \pi^2}\left\langle\int_{(h/\lambda)^2}^\infty \sum_{n=-\infty}^{\infty} \frac{e^{-\frac{n^2\beta^2}{4
\pt}}}{\mathcal{T}^3} \mathrm{d}\mathcal{T}\right\rangle.
\end{align}
The representation (\ref{A-PFA-4}) is suitable for zero and low temperatures,
whereas for high temperatures one should use in (\ref{A-PFA-4}) the Poisson
resummed winding sum (\ref{FT-SaP-6}).  We encounter cumulants of worldline
extents $\lambda$ in low and high temperature limits which can be determined via the
analytic expression \cite{Gies:2006cq}
\begin{align}\label{A-PFA-5}
\left\langle\lambda^D\right\rangle=D(D-1) \Gamma(D/2) \zeta(D).
\end{align}
Let us now point out the difference between the PFA and the worldline
approach. In the PFA, we always use one dimensional worldlines to determine
the distance, whereas the worldline dimension in the full formalism
corresponds to the dimension of the geometry. To obtain the Casimir force for
configurations containing one infinite plate in the worldline formalism, we
integrate over {this} infinite plate as in the plate-based
approach. However, the integration does not stop at the end of the
{second body, which in the present case is a} sphere {or}
cylinder. At arbitrary large distances, there are still worldlines which see
the sphere/cylinder, i.e., we have to integrate to infinity. We therefore
expect the leading-order PFA to reflect best the exact force laws. However,
the propertime support is not the same, and thus worldlines see an effective
height different from the one of the leading-order PFA, see
Fig. {\ref{A-F1} and \ref{A-F2}}.

The shape of the effective worldline height is roughly the same
for zero and high temperatures.  But at low temperature, the worldlines are
reweighted. Only worldlines for large propertimes contribute considerably and
thus worldlines at larger distances from the sphere become increasingly more important. Also
their inner structure comes into play. Using the 'plate based' PFA, we {ignore these  effects and}
take into account only the region below the
sphere/cylinder {with the same function $h_\mathrm{PB}$} ; hence, the result is expected to be too small.
 
In the following, we apply Eq.~(\ref{A-PFA-4}) (multiplied by
$\mathrm{d}h/\mathrm{d}a$ if necessary) to find the PFA expressions for the
sphere and cylinder above an infinite plate, respectively.

\subsection{Sphere above a plate}

\subsubsection{Leading-order PFA}

For the sphere, the evaluation of the leading-order PFA results in an
especially simple expression,
\begin{align}\label{A-PFA-S-1a}
-\frac{\mathrm{d}}{\mathrm{d}a}E^\mathrm{PFA}_{\mathrm{LO}}&=2\pi \,c_\mathrm{PP}\frac{\mathrm{d}}{\mathrm{d}a}\int_0^\infty \frac{r\,\mathrm{d}r}{\left(a+r^2/2 R\right)^3}
\\
&=2\pi R\,c_\mathrm{PP}\frac{\mathrm{d}}{\mathrm{d}a}\int_{a}^\infty \frac{\mathrm{d}h_\mathrm{LO}}{h_\mathrm{LO}^3}\notag=2\pi R\,\varepsilon^\mathrm{PP}(a).
\end{align}
Obviously, the relation $F^\mathrm{PFA}_{\mathrm{LO}}=2\pi
R\,\varepsilon^\mathrm{PP}(a)$ remains valid also at finite temperature. We
thus obtain
\begin{align}\label{A-PFA-S-1}
F^\mathrm{PFA}_{\mathrm{LO}}(a, T=0)=-\frac{2\pi R\,c_\mathrm{PP}}{a^3}=-\frac{\pi^3 R\,}{720 a^3}.
\end{align}
At finite $T$ and small $aT$ ($aT\lessapprox1/2$), Eq.~(\ref{A-PFA-4}) yields
\begin{align}
\Delta F^\mathrm{PFA}_{\mathrm{LO}}(a,T)=-\frac{ R\, \zeta(3)}{2} T^3+
\frac{ aR\, \pi^3 }{45} T^4. \label{A-PFA-S-2}
\end{align}  
For large $a T$ {($aT\gtrapprox1/2$)}, the expression (\ref{A-PFA-2b}) leads directly to 
\begin{align}
\Delta F^\mathrm{PFA}_{\mathrm{LO}}(a,T)
&=\frac{\pi^3 R\, }{720 a^3}-\frac{ R\, \zeta(3)   }{8 a^2} T\label{A-PFA-S-3a}
\\
&=-F^\mathrm{PFA}_{\mathrm{LO}}(a,0)+T \widetilde{F}^\mathrm{PFA}_{\mathrm{LO}}(a).\label{A-PFA-S-3}
\end{align}
Note that at $a=0$ the leading-order PFA predicts a $T^3$ behavior of the
thermal force for all $T$. At finite $a$, the validity of the low-temperature
limit is independent of $R$. With increasing $a$ the absolute value of the
PFA thermal force is always reduced, irrespective of $T$, quite the contrary to
the full worldline results as discussed in the main text.

\subsubsection{Plate-based PFA}
 Using Eq.~(\ref{A-PFA-4}), we obtain
\begin{widetext}
\begin{align}
-\Delta F^\mathrm{PFA}_{\mathrm{PB}}(a,T)&=
\frac{ 1 }{8 \pi}\int_0^R r \,
\mathrm{d}r\left\langle\sum_{n=1}^\infty\int_{h_{\mathrm{PB}}^2/\lambda^2}^\infty
  \frac{e^{-\frac{\beta^2 n^2}{4 \mathcal{T}}}}{\mathcal{T}^3}\,
  \mathrm{d}\mathcal{T}\right\rangle=\frac{\zeta(4) R^2}{ \pi\beta
  ^4}+\label{A-PFA-S-4} 
\\
&\!\!\!\!\!\!\!\!\!\!\!\!\!\!\!\!\!\!\!\!\!\!\!\!\!\!\!\!\!\!\!\!\!\!\!\!\!\frac{1}{
  \pi}\sum_{n=1}^\infty\left\langle\frac{ a e^{-\frac{n^2 \beta ^2 \lambda
        ^2}{4 a^2}} (a+2 R)}{n^4 \beta ^4}-\frac{ e^{-\frac{n^2 \beta ^2
        \lambda ^2}{4 (a+R)^2}} (a+R)^2}{n^4 \beta ^4}-\frac{ \sqrt{\pi }
    (a+R) \lambda  \text{Erfc}\left(\frac{n \beta  \lambda }{2 a}\right)}{2n^3
    \beta ^3}+\frac{ \sqrt{\pi } (a+R) \lambda  \text{Erfc}\left(\frac{n \beta
        \lambda }{2 (a+R)}\right)}{2n^3 \beta ^3}\right\rangle. \notag 
\end{align}
\end{widetext}
Let us first analyze Eq. (\ref{A-PFA-S-4}) for $a=0$,
\begin{align}
\Delta F^\mathrm{PFA}_{\mathrm{PB}}(a=0,T)&=-\frac{\zeta(4) R^2}{ \pi}T^4+\label{A-PFA-S-5}
\\
&\!\!\!\!\!\!\!\!\!\!\!\!\!\!\!\!\!\!\!\!\!\!\!\!\!\!\!\!\!\!\!\!\!\!\!\!\!\!\!\!\!\!\frac{1 }{ \pi}\sum_{n=1}^\infty\left\langle \frac{ e^{-\frac{n^2 \lambda ^2}{4 T^2 R^2}} R^2}{n^4 }T^4-\frac{ \sqrt{\pi } R \lambda  \text{Erfc}\left(\frac{n  \lambda }{2 T R}\right)}{2n^3 }T^3\right\rangle.\notag
\end{align}
Equation (\ref{A-PFA-S-5}) distinguishes low- ($T{\ll}1/R$) and
high-temperature ($T{\gg}1/R$) regimes. {The low-temperature regime is
  already well approached for $T\lessapprox1/2R$. For higher $T$, the thermal
  force is in the high-temperature regime, $T\gtrapprox1/2R$. }

At low temperatures, we have a $T^4$ behavior which is given
by the first term in Eq. (\ref{A-PFA-S-5}).  For {higher} $T$, this $T^4$ term is
canceled by the $T^4$ term {with} the exponential function, such that the leading
behavior is given by the $T^3$ term. Then, expanding Eq. (\ref{A-PFA-S-5}),
we get a $T^2$ contribution:
\begin{align}
\Delta F^\mathrm{PFA}_{\mathrm{PB}}(a=0,T)&=-\frac{  \zeta(3) R }{2 }T^3+\frac{\zeta(2) }{ 4\pi}\left\langle  \lambda^2 \right\rangle T^2 \notag
\\&=-\frac{  \zeta(3) R }{2 }T^3+\frac{\zeta(2)\zeta(2) }{ 2\pi} T^2.
\label{A-PFA-S-6}
\end{align}
Subtracting Eq.~(\ref{A-PFA-S-6}) from Eq.~(\ref{A-PFA-S-5}) and performing
the Poisson resummation, we obtain the full $T{\gtrapprox}1/2R$ behavior { at $a=0$}
\begin{align}
\Delta F^\mathrm{PFA}_{\mathrm{PB}}(0,T)=-\frac{\zeta(3) R T^3}{2}&+\frac{\pi^3 T^2 }{ 72} \notag
\\ 
&- \frac{\zeta(3)T}{  8 R}+\frac{\pi^3}{1440 R^2}.\label{A-PFA-S-7}
\end{align}
Thus, the leading large-$T$ behavior at $a=0$ is $\sim T^3$.
 
Let us now consider the case $a\neq 0$.  For $a\ll R$ and low temperature
${T\lessapprox1/2(R+a)\approx 1/2 R}$, we have a $T^4$ behavior given by the first term in
Eq.~(\ref{A-PFA-S-4}).  The dependence on $a$ is exponentially suppressed. This
corresponds to the case of two parallel plates with an area of $\pi R^2$, where
the dependence on $a$ is suppressed exponentially as well.

At medium temperature, ${2(R+a)}\approx {2}R\gtrapprox{1/T}\gtrapprox{2}a$ only the second and third term in
Eq.~(\ref{A-PFA-S-4}) are exponentially suppressed and can be neglected. The
leading {order} can be found by expanding the remainder {and considering only
  the converging sums}. 

To find the subleading terms, we again perform the Poisson
resummation. For medium temperature ${2}R\gtrapprox{1/T}\gtrapprox{2}a$ and $a\ll R$, we then obtain
\begin{align}
\Delta F^\mathrm{PFA}_{\mathrm{PB}}(a,T)&=\frac{a (a+2 R)\pi^3 T^4}{90}-\frac{(a+R)  \zeta(3) T^3}{2}\label{A-PFA-S-9}
\\
&+\frac{ \pi^3 T^2}{72  }-\frac{\zeta(3)}{8 (a+R)}T+\frac{\pi ^3}{1440 (a+R)^2}.\notag
\end{align}
The high-temperature limit {for}  ${1/T}\lessapprox{2}a$ can be performed irrespective of the
actual value $a/R$  by summing up the whole
Eq.~(\ref{A-PFA-S-4}). The result reads
\begin{align}
\Delta F^\mathrm{PFA}_{\mathrm{PB}}(a,T)
&=\frac{\pi ^3 R^2 (3 a+2 R)}{1440 a^3 (a+R)^2}-\frac{R^2\text{  }\zeta(3)}{8 a^2 (a+R)}T \label{A-PFA-S-10}
\\
&=-F^\mathrm{PFA}_{\mathrm{PB}}(a,0)+T \widetilde{F}^\mathrm{PFA}_{\mathrm{PB}}(a).\label{A-PFA-S-11}
\end{align}
Note that Eq.~(\ref{A-PFA-S-10}) reduces to  Eq.~(\ref{A-PFA-S-3a}) for
$a\rightarrow 0$ as it should. 

{For $a$ larger than $a\approx R$}, the following temperature behavior
occurs. At low temperature ${1/T}\gtrapprox{2}(R+a)$, a $T^4$
behavior arises from the first term in Eq.~(\ref{A-PFA-S-4}). At higher
temperatures, the behavior becomes rapidly linear as given by
Eq.~(\ref{A-PFA-S-9}), being valid for ${1/T}\lessapprox{2}a$.

The plate-based force can be obtained  in closed form {from Eq. (\ref{A-PFA-2b})} also without using
  the worldline language:
\begin{widetext}
\begin{align}
\Delta F^\mathrm{PFA}_{\mathrm{PB}}(a,T)&=\frac{\pi^3 R^2 (3 a+2 R)}{1440 a^3 (a+R)^2}\notag
\\
&-\frac{ T}{8}\sum _{n=1}^\infty  \Big[\frac{\text{coth}(2 n \pi  (a+R) T)}{(a+R)n^3 } +\frac{\text{csch}^2(2 a n \pi  T) \left[2 a n \pi  R T+\left(R-(a+R)/2\right) \text{sinh}(4 a n \pi  T)\right]}{ a^2 n^3}\Big].
\label{A-PFA-S-12}
\end{align}

\subsubsection{Sphere-based PFA}

For the sphere-based PFA, the thermal Casimir force is given by
\begin{equation}\label{A-PFA-S-13}
\Delta F_\mathrm{SB}^{\mathrm{PFA}}(a,T)=-\left\langle\frac{R^2}{8\pi}  \int_0^{\pi/2} \sin(\theta)h_\mathrm{SB}'(a)\mathrm{d}\theta 
\int_{h_\mathrm{SB}^2/\lambda^2}^\infty \sum_{n=1}^\infty \frac{ e^{-\frac{n^2 \beta^2}{4 \mathcal{T}}}}{\mathcal{T}^3} \,  \, \mathrm{d}\mathcal{T}\right\rangle,
\end{equation}
where $h_\mathrm{SB}(a)$ is given by (\ref{A-PFA-3b}).
For $a=0$, we obtain {the PFA approximation using the worldline language}
\begin{align}\label{A-PFA-S-14}
\Delta F_\mathrm{SB}^{\mathrm{PFA}}(a=0,T)=&\left\langle\frac{2 R^2\text{ln}\left(\frac{2 R T}{n \lambda}\right)}{n^4 \pi}T^4-\frac{\gamma R^2 \zeta(4)}{ \pi}T^4+\frac{R  \zeta(3)}{2 }T^3\right.\notag
\\
&\left.-\sum_{n=1}^\infty\frac{R^2 T^4}{4 n^4 \pi} \left(4+\frac{ n^2 \lambda^2}{R^2 T^2}\right) \exp\left(-\frac{ n^2 \lambda^2}{4 T^2 R^2}\right) \left(\pi  \text{Erfi}\left(\frac{ n \lambda}{2 R T}\right)-\text{Ei}\left(\frac{ n^2 \lambda^2}{4 R^2 T^2}\right)\right)\right\rangle,
\end{align}
\end{widetext}
where $\gamma$ is Euler's constant. The expansion in $T$ does not terminate
after a few terms, so we concentrate on the two leading coefficients. The
coefficient in front of $T^4$ contains the worldline average $\langle\ln
\lambda\rangle$. For an analytical expression, we note that
$\ln\lambda=m\ln\lambda^{1/m}$. For large $m$, we get $\lambda^{1/m}\rightarrow
1$, such that we can expand the logarithm,
\begin{align}\label{A-PFA-S-15}
\langle\ln \lambda\rangle=\langle\lim_{m\rightarrow \infty}m (\lambda^{1/m}-1)\rangle=-1-\gamma/2+\ln(2\pi),
\end{align}
where we have used Eq. (\ref{A-PFA-5}). Thus, the small-$T$ limit of
Eq. (\ref{A-PFA-S-14}) reads
\begin{align}\label{A-PFA-S-16}
\Delta F_\mathrm{SB}^{\mathrm{PFA}}(0,T)&= \frac{R^2 (2 \zeta'(4)+\zeta(4) (3+2\ln\left( \frac{R T}{\pi}\right)))}{ \pi} T^4 \notag
\\
&- R^3 \zeta(5)T^5.
\end{align}
At $a=0$, {the PFA estimate $|\Delta F_\mathrm{SB}^{\mathrm{PFA}}(a=0,T)|$} lies above {$|\Delta F_\mathrm{PB}^{\mathrm{PFA}}(a=0,T)|$, see Fig. \ref{FT-SaP-F1}}. For not
too small $T$, the {worldline result} lies above both {these}  PFA predictions, but due to the
logarithm in the $T^4$ coefficient the sphere-based PFA becomes larger at
smaller $T$, such that the {worldline} force enters the area spanned by the PFA prediction,
see Fig. \ref{FT-SaP-F1}.

The high-temperature limit can be obtained by expanding Eq. (\ref{A-PFA-S-14})
about $T=\infty$. The converging terms give the leading-order behavior. For
the subleading orders, one has to perform the Poisson summation, however, the
integral involved is rather complicated and {may still be inflicted with
  artificial convergence problems}. The leading-order behavior for {$a=0$ and large $T$} reads
\begin{align}\label{A-PFA-S-17}
\Delta F_\mathrm{SB}^{\mathrm{PFA}}(0,T)&= -\frac{R\zeta(3)}{2}T^3+\frac{\pi^3}{72}T^2+\mathcal{O}(T),
\end{align}
and corresponds to the leading behavior of the plate-based limit (\ref{A-PFA-S-7}). 

{Let us turn to the case of finite $a$. Expanding Eq. (\ref{A-PFA-S-13}),
  we obtain the $a$ dependent part of the thermal force,}
\begin{align}\label{A-PFA-S-18}
\Delta F_\mathrm{SB}^{\mathrm{PFA}}(a,T)-&\Delta F_\mathrm{SP}^{\mathrm{PFA}}(0,T)= \frac{2 a R \zeta(4) T^4}{\pi }\times\notag
\\
&\left(1-\frac{a}{2 R}+\frac{a^2}{3 R^2}-\frac{a^3}{4 R^3}+\dots\right)
\end{align}
The series has a form of $(R/a)\ln(1+a/R)$, {which we verified explicitly
  to 10th order.} 
Assuming that this form holds to all orders, we get
\begin{align}\label{A-PFA-S-19}
\Delta F_\mathrm{SB}^{\mathrm{PFA}}(a,T)-\Delta F_\mathrm{SP}^{\mathrm{PFA}}(0,T)&= \frac{ R^2 \pi^3 T^4}{45}\ln\left(1+\frac{a}{R}\right)
\end{align}
Note that the first two {terms agree with} the $T^4$ coefficient of the
plate-based formula (\ref{A-PFA-S-9}); {we also see that the absolute
  value of the thermal force decreases with increasing $a$. As
  Eq. (\ref{A-PFA-S-19}) was obtained by interchanging summation and
  integration, we cannot expect Eq.~(\ref{A-PFA-S-19}) to describe the full $a$
  dependence for all $a$ and $T$. Indeed at $a$
  fixed, the thermal correction $\Delta F_\mathrm{SB}^{\mathrm{PFA}}(a,T)$
  becomes $\sim T$ as $T\rightarrow \infty$, which is clearly not the
  case for Eq. (\ref{A-PFA-S-19}).
}

{We can estimate the range of applicability of Eq.~(\ref{A-PFA-S-19}) as
  follows. At high temperature and $a\approx0$, all PFA estimates agree. For
  large $T$, the leading behavior is $\sim T^3$, see
  e.g. Eq.~(\ref{A-PFA-S-17}). With increasing $a$ the force is still
  attractive. Demanding $\Delta F_\mathrm{SB}^{\mathrm{PFA}}(a,T)<0$ we see
  that that Eq.~(\ref{A-PFA-S-19}) leads to a positive thermal force for
  $a\gtrsim 1/T$. On the other hand in the low-temperature regime, the $a=0$
  contribution is given by Eq.~(\ref{A-PFA-S-16}). Taking only the leading
  contribution into account and demanding $\Delta
  F_\mathrm{SB}^{\mathrm{PFA}}(a,T)<0$, we again obtain that the force becomes
  positive at $a \gtrsim 1/T$. These rather rough estimates demonstrate
  that  the validity range for $a$ becomes
  narrower with increasing temperature. For very small $a$, however the thermal correction is linear in
  $a$ irrespectively of $T$, whereas the dependence
  on $a$ in the plate-based PFA is exponentially suppressed for small $T\ll 1/(R+a)$.}

At large temperatures $T>1/a$, we have the familiar situation
\begin{align}
\Delta F^\mathrm{PFA}_{\mathrm{SB}}(a,T)
=-F^\mathrm{PFA}_{\mathrm{SB}}(a,0)+T \widetilde{F}^\mathrm{PFA}_{\mathrm{SB}}(a),\label{A-PFA-S-20}
\end{align}
where 
\begin{align}
F^\mathrm{PFA}_{\mathrm{SB}}(a,0)=-\frac{\pi ^3 \left(6 a^2-3 a R+2 R^2\right)}{1440 a^3 R}-\frac{\pi ^3 \ln\left(\frac{a}{a+R}\right)}{240 R^2},\label{A-PFA-S-21}
\end{align}
and
\begin{align}
\widetilde{F}^\mathrm{PFA}_{\mathrm{SB}}(a)=\frac{(R-2 a) \zeta(3)}{8 a^2}-\frac{\ln \left(\frac{a}{a+R}\right) \zeta(3)}{4 R}.\label{A-PFA-S-22}
\end{align}

\subsection{Cylinder above a plate}

\subsubsection{Leading order PFA}
Unfortunately, {a simple relation similar to} $F^\mathrm{PFA}_{\mathrm{LO}}=2\pi
R\,\varepsilon^\mathrm{PP}(a)$ does not hold any longer for the cylinder, such
that the resulting formulas are not related to the known results of parallel
plates and are rather complicated. For arbitrary $a$ and $T$, we obtain
\begin{widetext}
\begin{align}\label{A-PFA-S-23}
  \frac{F^\mathrm{PFA}_{\mathrm{LO}}(a,
    T)}{L_y}=\sum_{n=1}^\infty\left\langle\frac{\lambda^2 \sqrt{a R}T^2}{4
      \sqrt{2} a^2 n^2 \pi}\left(_2
      F_2\left(\frac{3}{4},\frac{5}{4};1,\frac{3}{2};-\frac{\lambda^2 n^2 }{4
          a^2 T^2}\right)-\ _2
      F_2\left(\frac{3}{4},\frac{5}{4};\frac{3}{2},2;-\frac{\lambda^2 n^2 }{4
          a^2 T^2}\right)\right)\right\rangle,
\end{align}
\end{widetext}
where $_2 F_2$ is the hypergeometric function in the standard
notation. Eq. (\ref{A-PFA-S-23}) does not distinguish between $a<R$ and $a>R$,
since the relevant parameter for different temperature regions is $a T$. For
small $aT$ ($aT \lessapprox 1/2$), we can expand Eq.~(\ref{A-PFA-S-23}), resulting
in
\begin{align}\label{A-PFA-S-24}
\frac{F^\mathrm{PFA}_{\mathrm{LO}}(a, T)}{L_y}=&\frac{3 \sqrt{R}  \zeta(7/2)\zeta(1/2)}{4 \sqrt{2}  \pi }T^{7/2}
\\
&-\frac{15 a  \sqrt{R} \zeta(9/2)\zeta(-1/2)}{4 \sqrt{2}  \pi }T^{9/2}+\mathcal{O}(a^2). \notag
\end{align}
For large $a T$, the Poisson resummation of Eq.~(\ref{A-PFA-S-23}) leads  to 
%
\begin{align}
\Delta F^\mathrm{PFA}_{\mathrm{LO}}(a,T)
&=\frac{\pi ^3 \sqrt{a R}}{768 \sqrt{2} a^4}-\frac{3 \sqrt{a R} \zeta(3)}{32 \sqrt{2} a^3} T,\label{A-PFA-S-25}
\end{align}
which is, of course, $-F^\mathrm{PFA}_{\mathrm{LO}}(a,0)+T
\widetilde{F}^\mathrm{PFA}_{\mathrm{LO}}(a)$.  Note that at $a=0$ the
leading-order PFA predicts a $T^{7/2}$ behavior of the thermal force for all
$T$. At finite $a$, the validity of the low-temperature limit is independent of
$R$. With increasing $a$, the absolute value of the thermal force is always
reduced, irrespective of $T$, quite the contrary to the full worldline results.

\subsubsection{Plate-based PFA}

Here, we give only the analytic expressions for special limits, since no
general expression could be found in a closed form.  At $a=0$ and $T\ll 1/R$,
the thermal force can be found from the result of two parallel plates with an
area of $A=2 R L_y$,
\begin{align}
\Delta F^\mathrm{PFA}_{\mathrm{PB}}(a=0,T)=-2 R L_y\frac{\pi^2}{90} T^4.\label{A-PFA-S-26}
\end{align}
As temperature rises, the $T$ behavior changes from $T^4$ to $T^{7/2}$. For
$T\gg1/R$ and $a=0$, the plate-based PFA agrees with the leading-order PFA,
and the thermal force is given by the first term in Eq.~(\ref{A-PFA-S-24}). At
low temperatures and $a\ll R$, the dependence on $a$ is exponentially
suppressed, just as in the case of the plate-based PFA for the sphere.

Finally, at finite $a$ and $T\gg 1/a$, the force becomes classical
$-F^\mathrm{PFA}_{\mathrm{PB}}(a,0)+T
\widetilde{F}^\mathrm{PFA}_{\mathrm{PB}}(a)$, with
\begin{align}
\frac{F^\mathrm{PFA}_{\mathrm{PB}}(a,0)}{L_y}=&-\frac{(15+2 a (2+a) (11+3 a (2+a))) \pi ^2}{1440 a^3 (1+a)^2 (2+a)^3}\notag
\\
&-\frac{(5+4 a (2+a)) \pi ^3}{960 a^{7/2} (2+a)^{7/2}}\label{A-PFA-S-27}
\\
&-\frac{(5+4 a (2+a)) \pi ^2 \text{ArcTan}\left[\frac{1}{\sqrt{a (2+a)}}\right]}{480 a^{7/2} (2+a)^{7/2}},\notag
\end{align}
and 
\begin{align}
\frac{\widetilde{F}^\mathrm{PFA}_{\mathrm{PB}}(a)}{L_y}=&-\frac{3 (1+a) \zeta(3)}{16 a^{5/2} (2+a)^{5/2}}-\frac{\left(3+4 a+2 a^2\right) \zeta(3)}{8 a^2 (1+a) (2+a)^2 \pi } \notag
\\
&-\frac{3 (1+a) \text{ArcTan}\left[\frac{1}{\sqrt{a (2+a)}}\right] \zeta(3)}{8 a^{5/2} (2+a)^{5/2} \pi }.\label{A-PFA-S-28}
\end{align}
In Eqs. (\ref{A-PFA-S-27}) and (\ref{A-PFA-S-28}), we set $R=1$; general
expressions can be reconstructed by simple dimensional analysis.

\subsubsection{Cylinder-based PFA}

For the cylinder-based PFA, the thermal Casimir force is given by
\begin{align}\label{A-PFA-S-29a}
\Delta F_\mathrm{CB}^{\mathrm{PFA}}(a,T)&=-\frac{R L_y}{8\pi^2}  \int_0^{\pi/2} h_\mathrm{SB}'(a)\mathrm{d}\theta \notag
\\
&\times\int_{h_\mathrm{CB}^2/\lambda^2}^\infty \sum_{n=1}^\infty \frac{ e^{-\frac{n^2 \beta^2}{4 \mathcal{T}}}}{\mathcal{T}^3} \,  \, \mathrm{d}\mathcal{T},
\end{align}
where $h_\mathrm{CB}(a)$ is given by Eq.~(\ref{A-PFA-3b}).
At $a=0$, the thermal force can be found in closed form,
\begin{widetext}
\begin{align}
&\frac{\Delta F^\mathrm{PFA}_{\mathrm{CB}}(a=0,T)}{L_y}=\Big\langle\sum_{n=1}^\infty \frac{c}{1680}\sqrt{x } \Big[ 35 e^{-\frac{x ^2}{8}} \pi  x ^{3/2}\left(I_{-1/4}\left(\frac{x ^2}{8}\right)+7 I_{3/4}\left(\frac{x ^2}{8}\right)\right)\label{A-PFA-S-29}
\\
&-1260 \sqrt{2} \Gamma\left(\frac{3}{4}\right) \ _1 F_1\left(\frac{3}{4};\frac{1}{2};-\frac{x ^2}{4}\right)\notag
\\
&-16 x ^{3/2} \left(-35 \ _2F_2\left(1,\frac{3}{2};\frac{5}{4},\frac{7}{4};-\frac{x ^2}{4}\right)+16 \ _2F_2 \left(1,\frac{3}{2};\frac{9}{4},\frac{11}{4};-\frac{x ^2}{4}\right)\right.\notag
\\
&\left.
+16 \ _2 F_2\left(\frac{3}{2},2;\frac{9}{4},\frac{11}{4};-\frac{x ^2}{4}\right)+3 \ _3 F_3\left(1,1,\frac{3}{2};2,\frac{9}{4},\frac{11}{4};-\frac{x ^2}{4}\right)\right)\notag\notag
\\
&-504 \sqrt{2} x ^2 \Gamma \left(\frac{3}{4}\right) \ _2 F_2\left(\frac{5}{4},\frac{7}{4};\frac{3}{2},\frac{9}{4};-\frac{x ^2}{4}\right)+5 \sqrt{2} x ^3 \Gamma\left(-\frac{3}{4}\right) \ _2F_2\left(\frac{5}{4},\frac{7}{4};\frac{5}{2},\frac{11}{4};-\frac{x ^2}{4}\right)\Big]\Big\rangle,\notag
\end{align}
\end{widetext}
where $I_n$ is the modified Bessel function of the first kind, $_p F_q$ the
generalized hypergeometric function, $c=2 R T^4/n^4 \pi ^2$ and $x=n \lambda/2
R T$.  At small $T$, the expansion of Eq.~(\ref{A-PFA-S-29}) leads to
\begin{align}
\frac{\Delta F^\mathrm{PFA}_{\mathrm{CB}}(0,T)}{L_y}&=\frac{R T^4 \left(3 \pi ^4+2 \pi ^4 \ln\left(\frac{R T}{2 \pi }\right)+180 \zeta'(4)\right)}{90 \pi ^2}\notag
\\
&-\frac{R^2 T^5 \zeta(5)}{\pi }+\mathcal{O}(T^6).\label{A-PFA-S-30}
\end{align}
As temperature rises, the $T$ behavior changes to $T^{7/2}$. For $T\gg1/R$ and
$a=0$, the cylinder-based PFA agrees with the leading-order PFA and the
thermal force is given by the first term in Eq.~(\ref{A-PFA-S-24}).
For sufficiently small $a$  the difference to the $a=0$ result reads
\begin{align}
\frac{\Delta F^\mathrm{PFA}_{\mathrm{CB}}(a,T)-\Delta F^\mathrm{PFA}_{\mathrm{CB}}(0,T)}{L_y}&=\frac{a \pi^2}{45} T^4 +\mathcal{O}(T^6).\label{A-PFA-S-31}
\end{align}

At finite $a$ and $T\gg 1/a$, the force becomes classical
$-F^\mathrm{PFA}_{\mathrm{CB}}(a,0)+T
\widetilde{F}^\mathrm{PFA}_{\mathrm{CB}}(a)$, with
\begin{align}
\frac{F^\mathrm{PFA}_{\mathrm{CB}}(a,0)}{L_y}=&-\frac{\left(15+8 a+4 a^2\right) \pi ^2}{1440 a^3 (2+a)^3}\label{A-PFA-S-32}
\\ 
&-\frac{\sqrt{a (2+a)} \left(5+6 a+3 a^2\right) \pi ^3}{960 a^4 (2+a)^4} \notag
\\
&-\frac{\left(5+6 a+3 a^2\right) \pi ^2 \text{ArcTan}\left[\frac{1}{\sqrt{a (2+a)}}\right]}{480 a^{7/2} (2+a)^{7/2}},\notag
\end{align}
and 
\begin{align}
\frac{\widetilde{F}^\mathrm{PFA}_{\mathrm{PB}}(a)}{L_y}=&-\frac{3 R^2 \zeta(3)}{8 a^2 \pi  (a+2 R)^2} \label{A-PFA-S-33}
\\
&-\frac{R \left(a^2+2 a R+3 R^2\right) \zeta(3)}{16 (a (a+2 R))^{5/2}} \notag
\\
&\!\!\!\!\!\!\!\!\!\!\!\!\!\!\!\!\!\!\!\!-\frac{R \left(a^2+2 a R+3 R^2\right) \text{ArcTan}\left[\frac{R}{\sqrt{a (a+2 R)}}\right] \zeta(3)}{8 a^{5/2} \pi  (a+2 R)^{5/2}}. \notag
\end{align}
In Eqs.~(\ref{A-PFA-S-32}) and (\ref{A-PFA-S-33}), we set $R=1$, general
expressions can be reconstructed by dimensional analysis.

\acknowledgments

We thank T.~Emig for providing the data of Fig. \ref{CEff-RT0-2} and M.~Bordag
and E.~Elizalde for useful discussions. We have benefited from activities
within ESF Research network CASIMIR. AW acknowledges support by the
Landesgraduiertenf\"orderung Baden-W\"urttemberg, by the Heidelberg Graduate
School of Fundamental Physics. HG was supported by the DFG under contract
Gi 328/3-2 and Gi 328/5-1 (Heisenberg program).

\end{document}